\documentclass[superscriptaddress,twocolumn,pre,reprint,aps]{revtex4-2}
\usepackage{amsmath}
\usepackage{amssymb}
\usepackage{graphicx}
\usepackage{amsfonts}
\usepackage{bbm}
\usepackage{tikz}
\newcommand{\radialarrows}{
  \begin{tikzpicture}[scale=0.15, baseline={(0,-0.3)},>=stealth]
    \draw[<->, line width=0.4pt] (-1,0) -- (1,0);
    \draw[<->, line width=0.4pt] (0,-1) -- (0,1);
  \end{tikzpicture}
}

\newcommand{\hyponearrows}{
  \begin{tikzpicture}[scale=0.15, baseline={(0,-0.3)},>=stealth]
    \draw[<->, line width=0.4pt] (-1,0) -- (1,0);
    \draw[>-<, line width=0.4pt] (0,-1) -- (0,1);
  \end{tikzpicture}
}

\newcommand{\hyptwoarrows}{
  \begin{tikzpicture}[scale=0.15, baseline={(0,-0.3)},>=stealth]
    \draw[<->, line width=0.4pt] (-0.707,-0.707) -- (0.707,0.707);
    \draw[>-<, line width=0.4pt] (0.707,-0.707) -- (-0.707,0.707);
  \end{tikzpicture}
}

\usepackage{hyperref}
\begin{document}

\title{Heterogeneity-Induced Oscillations in Active Nematics}
\author{A.J.H. Houston}
\email{Alexander.Houston@glasgow.ac.uk}
\affiliation{University of Glasgow, School of Mathematics and Statistics, University Place, Glasgow G12 8QQ, United Kingdom.}
\author{M. Grinfeld}
\affiliation{University of Strathclyde, Department of Mathematics and Statistics, Livingstone Tower, 26 Richmond Street, Glasgow G1 1XH, United Kingdom.}
\author{G. McKay}
\affiliation{University of Strathclyde, Department of Mathematics and Statistics, Livingstone Tower, 26 Richmond Street, Glasgow G1 1XH, United Kingdom.}
\author{N.J. Mottram}
\affiliation{University of Glasgow, School of Mathematics and Statistics, University Place, Glasgow G12 8QQ, United Kingdom.}

\begin{abstract}
    One of the defining features of active nematics is that above a critical activity the quiescent state becomes unstable to a distorted, flowing one. We show that spatial variations in activity can fundamentally change the nature of this instability, affecting the symmetry of the unstable mode and producing spontaneous oscillations. We analytically identify a dynamical system for the evolution of the odd and even director modes, with the leading-order coefficients dependent on the activity profile, allowing a quantitative connection between the spatially-heterogeneous activity and dynamics, which we verify numerically. In the context of constant gradients in activity, we determine a phase diagram for the active response and highlight how variation of the activity profile causes the oscillations to vary from almost harmonic to relaxational. Our results indicate a novel route to spatio-temporal structure in active nematics and suggest experiments on controllable light-activated systems.
\end{abstract}

\maketitle
Active nematics are non-equilibrium systems in which the orientationally-ordered constituents convert energy into stresses \cite{ramaswamy2010mechanics,marchetti2013hydrodynamics,doostmohammadi2018active}. The addition of such active stresses into the hydrodynamic theories of passive nematics \cite{de1993physics} has enabled the modelling of cell monolayers \cite{duclos2017topological}, tissues \cite{saw2017topological}, bacteria \cite{zhou2014living} and synthetic microtubule suspensions \cite{sanchez2012spontaneous}. These active nematics are distinguished from their passive counterparts by the fundamental hydrodynamic instability of the  undistorted, quiescent  state to a distorted, flowing one \cite{aditi2002hydrodynamic,ramaswamy2010mechanics}. Such collective flows are connected to the spontaneous dynamics of defects \cite{giomi2014defect,duclos2020topological,binysh2020three,houston2022defect} and colloids \cite{loewe2022passive,houston2023active,houston2023colloids,ray2023rectified,houston2025role}, having consequences for both biological functionality \cite{saw2017topological,maroudas2021topological,copenhagen2021topological} and the extraction of work from active systems \cite{di2010bacterial}. In confined systems, this spontaneous flow transition occurs at a critical level of activity, determined by the competition between active and elastic stresses. This flow transition precedes the development of defect dynamics \cite{shendruk2017dancing} and ultimately active turbulence \cite{alert2022active}, that occur at higher levels of activity. Since the original theoretical prediction \cite{voituriez2005spontaneous}, the transition has been confirmed both numerically \cite{marenduzzo2007steady} and experimentally \cite{duclos2018spontaneous}, and has spawned many subsequent studies \cite{keogh2022helical,mori2023viscoelastic,alam2024active,vaidya2024active,pratley2024three,houston2024spontaneous,lavi2025nonlinear,bell2025ordering,aramini2025spontaneous}.

However, the current theoretical understanding of active nematics largely rests on the assumption of uniform activity. 
This assumption restricts our ability both  to   understand the consequences of inherent spatial heterogenity in biological systems and to exploit engineered heterogeneities. The former is near ubiquitous, for example due to distinct populations of bacteria or cells, variations within a population or variations in nutrients, while the latter has been demonstrated through light-activated systems \cite{zhang2021spatiotemporal}. While heterogeneous activity has been shown to allow spatiotemporal control of defects \cite{zhang2021spatiotemporal,shankar2024design,ruske2022activity} and turbulence \cite{partovifard2024controlling,schimming2025turbulence}, its consequences for the spontaneous flow transition has received only limited attention \cite{houston2024spontaneous}. 
Here we investigate the effect of spatially-varying activity on the spontaneous flow transition in flow-aligning active nematics, finding that it can lead to a variety of qualitatively distinct dynamics. The classical, spatially-uniform, activity instability proceeds via a double pitchfork bifurcation, in which two linearly independent modes become unstable. We show that spatially-varying activity can allow any linear combination of these modes to be selected as the fastest growing distortion. More strikingly, the undistorted quiescent state can lose stability via a Hopf bifurcation, leading to sustained oscillations. For constant activity gradients, we quantify how these oscillations can vary from almost harmonic to relaxational.

\begin{figure*}
    \centering
    \begin{tikzpicture}[scale=1.6,>=stealth]
        \node[anchor=south west,inner sep=0] at (0,0)
{\includegraphics[width=1\linewidth, trim = 50 423 20 220, clip, angle = 0, origin = c]{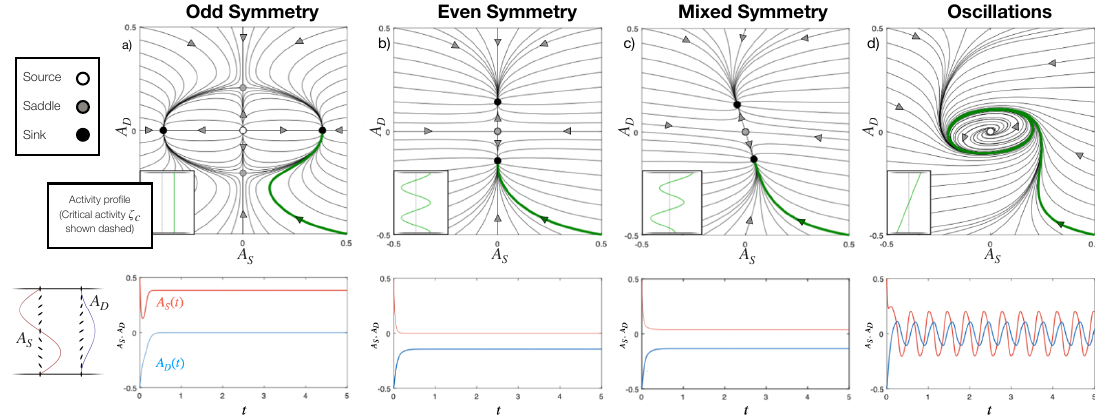}};
    \end{tikzpicture}
    \caption[The behaviour of a confined active nematic for different activity profiles.]{The behaviour of a confined active nematic for different activity profiles. The panels are organised according to the nature of the long-time director state, with the top row showing phase portraits for the activity profile shown in the inset, while the bottom row shows the variation of $A_S$ and $A_D$, the amplitudes of the odd and even symmetry modes, along the green highlighted trajectory. The evolution of the system is determined by numerical integration of the nonlinear director dynamics \eqref{eq:DirectorDynamics}.
    a) For a uniform activity above the instability threshold both modes are unstable at the origin, with $A_S$ having a higher growth rate and ultimately dominating the system. b) A perturbation of the form $\cos\left(4\pi z/d\right)$ produces a saddle point at the origin, with the instability leading to the $D$ mode dominating asymptotically. c) Adding a linear contribution to the perturbation in b) rotates both eigenvectors such that the system tends to a mixture of the $S$ and $D$ modes. d) A constant gradient centred on the critical activity results in a limit cycle and oscillations in the director modes. The particular non-zero values of the activity Fourier coefficients, defined in \eqref{eq:ActivityFourier}, are a): $a_0=1/2$, b): $a_2=1/2$, c): $a_2=1/2$, $b_1=-3/(4\pi)$, $b_2=-3/(8\pi)$ and d): $b_1=-3/\pi$, $b_2=-3/(2\pi)$.
    }
    \label{fig:PhasePortraits}
\end{figure*}

\textit{Theoretical model.} --- Our approach is to use the Ericksen--Leslie equations for active nemato-hydrodynamics, describing the dynamics of the director field $\mathbf{n}$, the average orientation of the active agents, and the flow velocity $\mathbf{u}$. We consider a channel geometry, with solid surfaces at $z=0$ and $z=d$ and assume that the director field remains in a plane, which we define as the $xz$-plane, with translational invariance along the $x$- and $y$-directions, so that $\mathbf{n}=\sin\theta(z)\mathbf{e}_x+\cos\theta(z)\mathbf{e}_z$. At both surfaces we impose homeotropic anchoring, corresponding to $\theta=0$, along with a no-slip condition for the flow. The dynamics of the director may be found, as detailed in Appendix \ref{AppA}, by utilising the incompressibility of the flow, solving for the flow velocity and then projecting  along a direction perpendicular to $\mathbf{n}$, yielding \begin{equation}
     \partial_t\theta=\frac{K}{\gamma}\partial^2_z\theta+\frac{1-\nu\cos2\theta}{2\mu}\left[\overline{\sigma_{xz}}-\sigma_{xz}(z)\right],
     \label{eq:DirectorDynamics}
\end{equation}
with $\overline{\bullet}$ denoting the average over the channel, and where $K$ is the isotropic elastic constant, $\gamma$ the rotational viscosity, $\nu$ the flow alignment parameter and $\mu$ the Newtonian viscosity. The stress tensor component, $\sigma_{xz}$, is given by
\begin{equation}
    \sigma_{xz}=-\dfrac{K}{2}\left(1-\nu\cos2\theta\right)\partial^2_z\theta-\frac{\zeta}{2}\sin2\theta,
    \label{eq:StressTensorxz}
\end{equation}
where $\zeta$ is the activity, which here we consider to be a spatially-varying function, 
with positive and negative values corresponding to extensile and contractile active stresses respectively. 

For homogeneous activity, linearising \eqref{eq:DirectorDynamics} and \eqref{eq:StressTensorxz} around the uniformly aligned ($\theta\equiv 0$) state reveals that two modes become simultaneously unstable at the critical activity, $\zeta_c$ \cite{voituriez2005spontaneous}, the expression for which is given in \eqref{eq:Zetac}. Of these modes, one has odd symmetry about the middle of the channel, $\theta_S=\sin\left(2\pi z/d\right)$, the other even, $\theta_D=\frac{1}{2}\left[1-\cos\left(2\pi z/d\right)\right]$, with the odd mode having the faster growth rate \cite{pratley2024three}. On account of their shape, and following \cite{pratley2024three}, we shall term them the $S$ and $D$ mode, respectively.

Figures \ref{fig:PhasePortraits} and \ref{fig:ActivityGradient} show numerical solutions to \eqref{eq:DirectorDynamics}. For these figures we fix $\gamma=\mu$, take $\nu=-1.1$ and scale by the elastic timescale $t_e=\gamma d^2/K$, this providing a more robust reference than an active timescale when making comparisons between different activity profiles. Figure \ref{fig:ActivityGradient}a) also makes a comparison with the following linear analysis, as well as the further analysis provided in the appendices.

\textit{Linear analysis.} --- 
Considering a spatially-varying perturbation about the critical activity, 
$\zeta(z)=\zeta_c(1+\varepsilon\Tilde{\zeta}(z))$, where $\varepsilon$ is a small parameter, we write the director field in terms of the $S$ and $D$ modes,  $\theta=A_S\theta_S+A_D\theta_D$, and construct a dynamical system in terms of the amplitudes, $A_S$ and $A_D$. Taking a Fourier series for the activity perturbation, 
\begin{equation}
    \Tilde{\zeta}=a_0+\sum_{n=1}^{\infty}\left[a_n\cos\left(\frac{2\pi nz}{d}\right)+b_n\sin\left(\frac{2\pi nz}{d}\right)\right],
    \label{eq:ActivityFourier}
\end{equation}
we have, at linear order, $\dot{\mathbf{A}}=\varepsilon\mathbf{\Lambda}\mathbf{A}$, or explicitly
\begin{equation}
    \begin{pmatrix}
        \dot{A}_S \\[1pt]
        \dot{A}_D
    \end{pmatrix}
    =
    \varepsilon\begin{pmatrix}
        \Lambda^S_S & \Lambda^S_D \\[3pt]
        \Lambda^D_S & \Lambda^D_D
    \end{pmatrix}
    \begin{pmatrix}
        A_S \\[3pt]
        A_D
    \end{pmatrix},
    \label{eq:DynamSysLin}
\end{equation}
with
\begin{align}
    \Lambda^S_S&=\frac{2\pi^2(1+\Gamma)}{t_e}(2a_0-a_2),
    \label{eq:LamdaSS}\\
    \Lambda^S_D&=\frac{\pi^2(1+\Gamma)}{t_e}(2b_1-b_2),
    \label{eq:LamdaSD}\\
    \Lambda^D_S&=-\frac{1}{3+2\Gamma}\frac{4\pi^2(1+\Gamma)}{t_e}b_2,
    \label{eq:LamdaDS}\\
    \Lambda^D_D&=\frac{1}{3+2\Gamma}\frac{4\pi^2(1+\Gamma)}{t_e}\left(a_0-a_1+\frac{1}{2}a_2\right),
    \label{eq:LamdaDD}
\end{align}
and $\Gamma=\gamma(1-\nu)^2/(4\mu)$. 

While the four components of $\boldsymbol{\Lambda}$ are linearly independent, allowing any linear dynamical system to be realised, the nature of the stability of the undistorted quiescent state, hereafter the trivial state, is determined by just $\text{Tr}\boldsymbol{\Lambda}$ and $\text{det}\boldsymbol{\Lambda}$ \cite{jordan2007nonlinear}. The two remaining degrees of freedom set the directions of the eigenvectors of $\boldsymbol{\Lambda}$. It is useful to treat separately the cases where $\Delta=4\,\text{det}\boldsymbol{\Lambda}-\left(\text{Tr}\boldsymbol{\Lambda}\right)^2$ is positive and negative \footnote{Note that we have taken $\Delta$ with the opposite sign convention than typically used. This is done to simplify the approximate oscillation frequency, given in \eqref{eq:GeneralFrequencyApproximation}.} and below we consider four example activity profiles to highlight the variety of qualitatively distinct dynamics that can occur (as shown in Figure \ref{fig:PhasePortraits}).

For $\Delta<0$: if $\text{det}\boldsymbol{\Lambda}>0$, the fixed point corresponding to the trivial state is a sink if $\text{Tr}\boldsymbol{\Lambda}<0$ and a source if $\text{Tr}\boldsymbol{\Lambda}>0$; if $\text{det}\boldsymbol{\Lambda}<0$ the  trivial state corresponds to a saddle regardless of the value of $\text{Tr}\boldsymbol{\Lambda}$. In the sink case the trivial state is therefore stable, as applies for uniform activity below the critical value $\zeta_c$. The source case is illustrated in Figure \ref{fig:PhasePortraits}a) for a uniform activity perturbation, corresponding to the classic instability \cite{voituriez2005spontaneous}. Here both modes are unstable but the $S$ mode has the larger growth rate. Figure \ref{fig:PhasePortraits}b) shows the saddle arising from a $\cos\left(4\pi z/d\right)$ deviation in activity from the critical value, leading to an instability through the $D$ mode. More generally, when the activity profile is even about the centre of the channel (so that all the $b_i$ are zero), $\boldsymbol{\Lambda}$ is diagonal and the eigenvectors are along the $S$ and $D$ modes. However, an instability through any linear combination of these modes can be promoted by adding odd components to the activity profile. This situation is illustrated in Figure \ref{fig:PhasePortraits}c), where the addition of a constant activity gradient to the profile in Figure \ref{fig:PhasePortraits}b) distorts the saddle so that growth occurs through a mixture of the $S$ and $D$ modes.

Conversely, when $\Delta>0$ the trivial state corresponds to a spiral sink ($\text{Tr}\boldsymbol{\Lambda}<0$), centre ($\text{Tr}\boldsymbol{\Lambda}=0$) or spiral source ($\text{Tr}\boldsymbol{\Lambda}>0$). The last of these is the most interesting and is illustrated in Figure \ref{fig:PhasePortraits}d) for a pure constant activity gradient, where, as discussed further in the next section, the linear terms can combine with stabilising nonlinear terms, leading to a limit cycle and oscillatory dynamics. In Figure \ref{fig:PhasePortraits}, panels b)-d) have been arranged so as to vary from a purely even to a purely odd deviation of the activity from the critical value, showing that odd activity components are necessary but not sufficient for oscillations.

The preceding reasoning only governs the dynamics of the active nematic when the distortion amplitudes are small. Determining the long-time evolution requires nonlinear terms. Nonetheless, the behaviour shown in Figure 1, obtained by numerical integration of \eqref{eq:DirectorDynamics}, accords with the expectation built from considering the dynamical system in \eqref{eq:DynamSysLin}.

\begin{figure}
    \centering
    \begin{tikzpicture}[scale=1.6,>=stealth]
        \node[anchor=south west,inner sep=0] at (0,0)
{\includegraphics[width=0.4\linewidth, trim = 0 0 0 0, clip, angle = 90, origin = c]{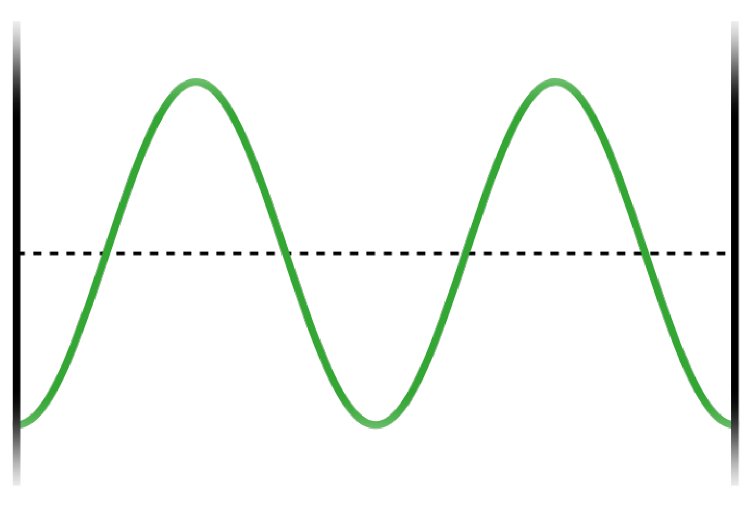}};
        \node[anchor=south west,inner sep=0] at (1.75,0)
{\includegraphics[width=0.4\linewidth, trim = 0 0 0 0, clip, angle = 90, origin = c]{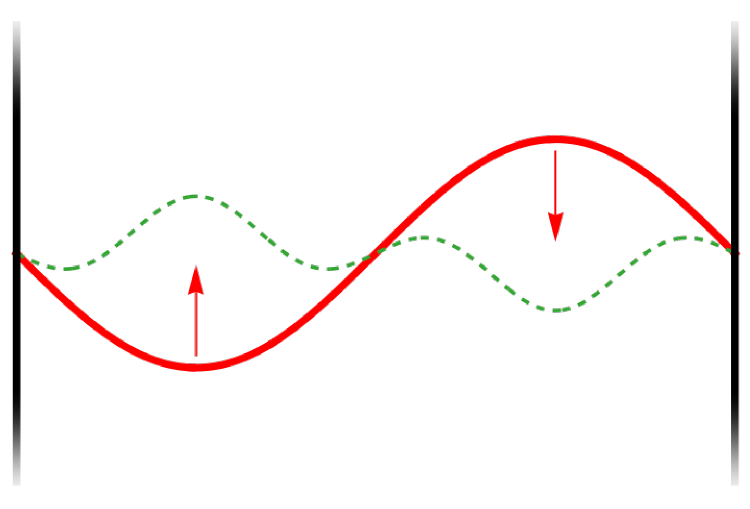}};
         \node[anchor=south west,inner sep=0] at (3.5,0)
{\includegraphics[width=0.4\linewidth, trim = 0 0 0 0, clip, angle = 90, origin = c]{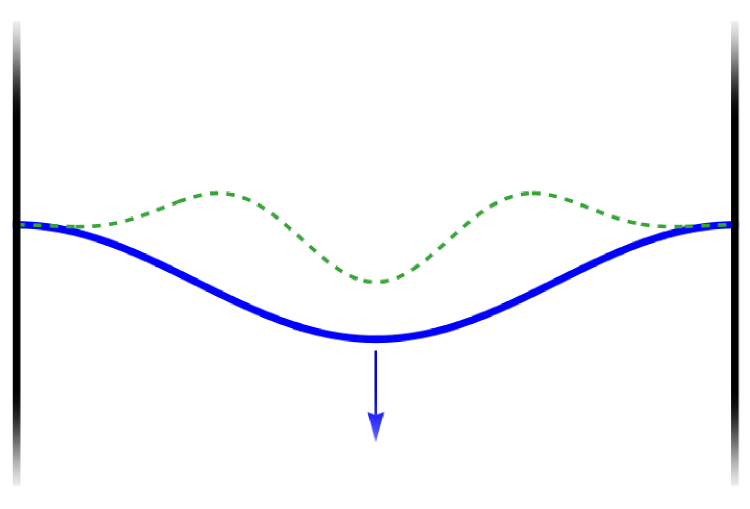}};

        \node[anchor=south west,inner sep=0] at (0,-2.5)
{\includegraphics[width=0.4\linewidth, trim = 0 0 0 0, clip, angle = 90, origin = c]{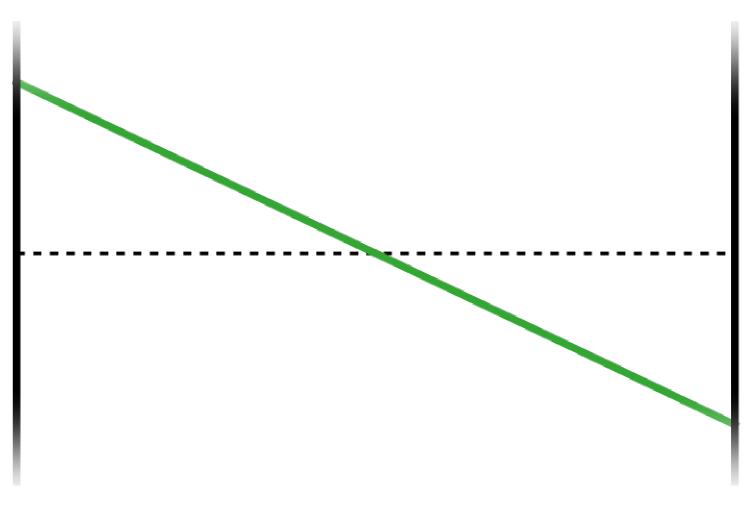}};
        \node[anchor=south west,inner sep=0] at (1.75,-2.5)
{\includegraphics[width=0.4\linewidth, trim = 0 0 0 0, clip, angle = 90, origin = c]{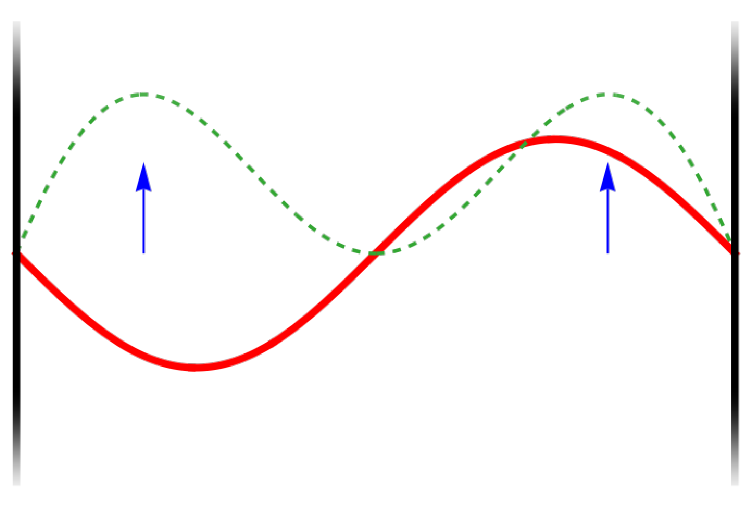}};
         \node[anchor=south west,inner sep=0] at (3.5,-2.5)
{\includegraphics[width=0.4\linewidth, trim = 0 0 0 0, clip, angle = 90, origin = c]{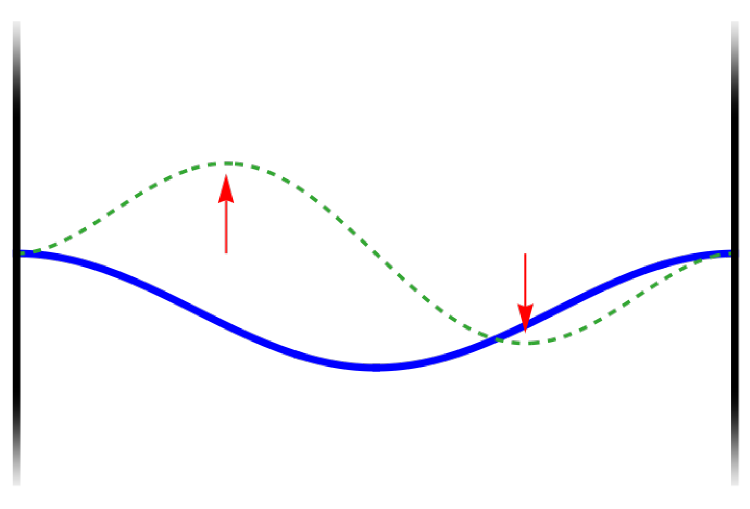}};

    \node[scale=0.8] at (0.75,2.4) {\textbf{Activity profile}};
    \node[scale=0.8] at (2.5,2.4) {\textbf{Effect on $S$ mode}};
    \node[scale=0.8] at (4.25,2.4) {\textbf{Effect on $D$ mode}};

    \node[scale=0.7,draw] at (2.89,1.25) {$S$ shrinks};
    \node[scale=0.7,draw] at (4.75,1.65) {$D$ grows};

    \node[scale=0.7,draw] at (2.89,-1.25) {$S\to D$};
    \node[scale=0.7,draw] at (4.75,-2.1) {$D\to-S$};
        
    \end{tikzpicture}
    \caption[The effect of activity profiles on director modes.]{ The effect of activity profiles on director modes. As described in the main text, the role of $\tilde\zeta$, the perturbation of the activity from the critical value, (denoted by the dashed line in the left-hand plots), on the dynamics of the director, $\theta$, can be inferred from their product, that is $\partial_t\theta\sim\Tilde\zeta\theta$. This product is illustrated by the green dashed line for both the $S$ and $D$ director modes and for two examples: $\Tilde{\zeta}\sim\cos\left(4\pi z/d\right)$ (top row) and $\Tilde{\zeta}\sim z-d/2$ (bottom row), these being the profiles in Figure \ref{fig:PhasePortraits}b) and d) respectively. The graphical reasoning accords with what is seen in Figure \ref{fig:PhasePortraits}: the top activity profile damps the $S$ mode and excites the $D$ mode, while the bottom acts in the way that can induce oscillations. Arrows indicate the induced dynamics of the director mode, with the colour indicating the symmetry of $\tilde\zeta\theta$.
    }
    \label{fig:GraphicalArgumentForActivityPatterns}
\end{figure}
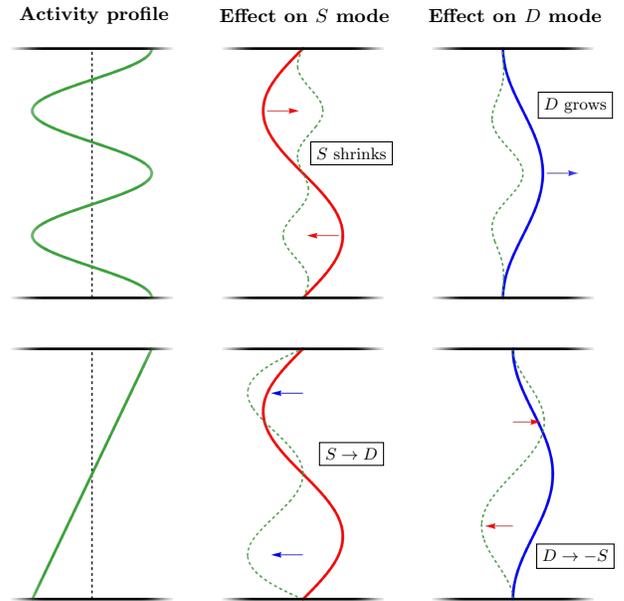

\begin{figure*}
    \centering
    \begin{tikzpicture}[scale=1.6,>=stealth]
        \node[anchor=south west,inner sep=0] at (0,0)
{\includegraphics[width=0.6\linewidth, trim = 0 0 0 0, clip, angle = 0, origin = c]{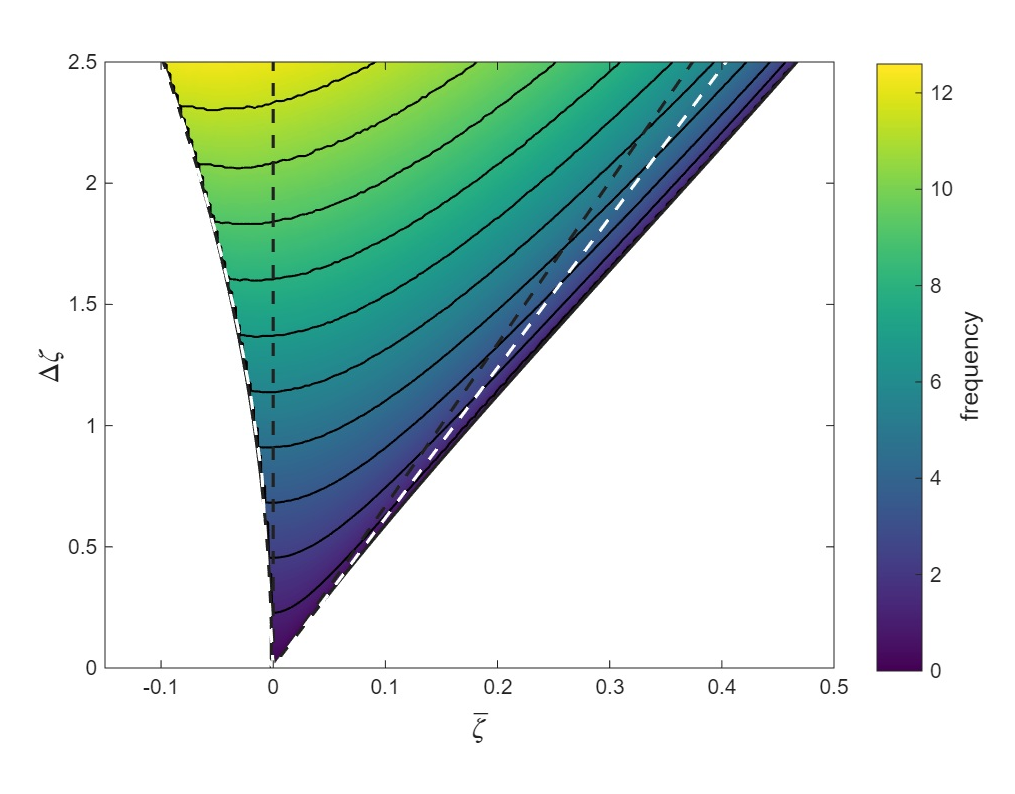}};

    \node[anchor=south west,inner sep=0] at (6.7,0.1)
{\includegraphics[width=0.385\linewidth, trim = 100 275 90 50, clip, angle = 0, origin = c]{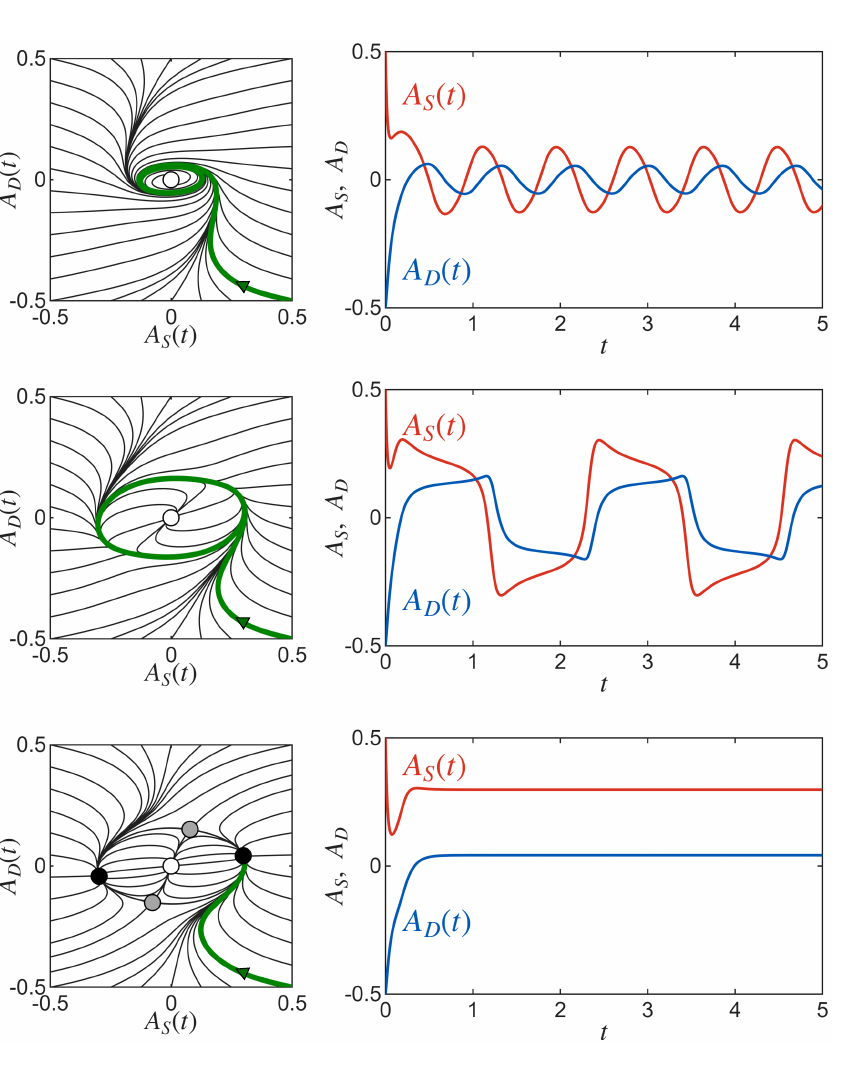}};
    
    \node[scale=1,draw,fill=white] at (1.82,3.25) {b)};
    \node[scale=1,draw,fill=white] at (3.7,3.25) {c)};
    \node[scale=1,draw,fill=white] at (3.7,1.65) {d)};

    \node[scale=1,fill=white] at (0.3,4.75) {a)};
    \node[scale=1,] at (6.65,5.25) {b)};
    \node[scale=1] at (6.65,3.55) {c)};
    \node[scale=1] at (6.65,1.75) {d)};

    \node[scale=1,draw,fill=white] at (0.95,1.65) {Trivial state};
    \node[scale=1,draw,fill=white] at (3,4.6) {Oscillations};
    \node[scale=1,draw,fill=white,align=center] at (4.85,1.3) {Non-trivial\\ steady state};
    
    \end{tikzpicture}
    \caption[The instabilities associated with the spontaneous flow transition for a homogeneous active nematic film.]{The instabilities of active nematics due to constant activity gradients. The activity profile is parameterised by $\bar{\zeta}$ and $\Delta\zeta$, characterising a shift of the average activity from its critical homogeneous value and the difference in activity across the channel, respectively, both of which are normalised by $\zeta_c$. a) Varying the activity parameters produces three regimes of behaviour: the trivial undistorted quiescent state to the left, a non-trivial steady state to the right and sustained oscillations in the middle. The solid boundaries are derived from numerical analysis of \eqref{eq:DirectorDynamics}, the black dashed approximations from linear analysis, and the left- and right-hand white dashed lines from the analysis of Appendix \ref{AppC} and \ref{AppD} respectively. b)-d) Phase portraits of the mode amplitudes $A_S$ and $A_D$, together with their evolution along the green highlighted trajectory, for the corresponding points in a). The dynamics show b) close-to-harmonic oscillations, c) relaxation oscillations and d) a non-trivial steady state.
    }
    \label{fig:ActivityGradient}
\end{figure*}
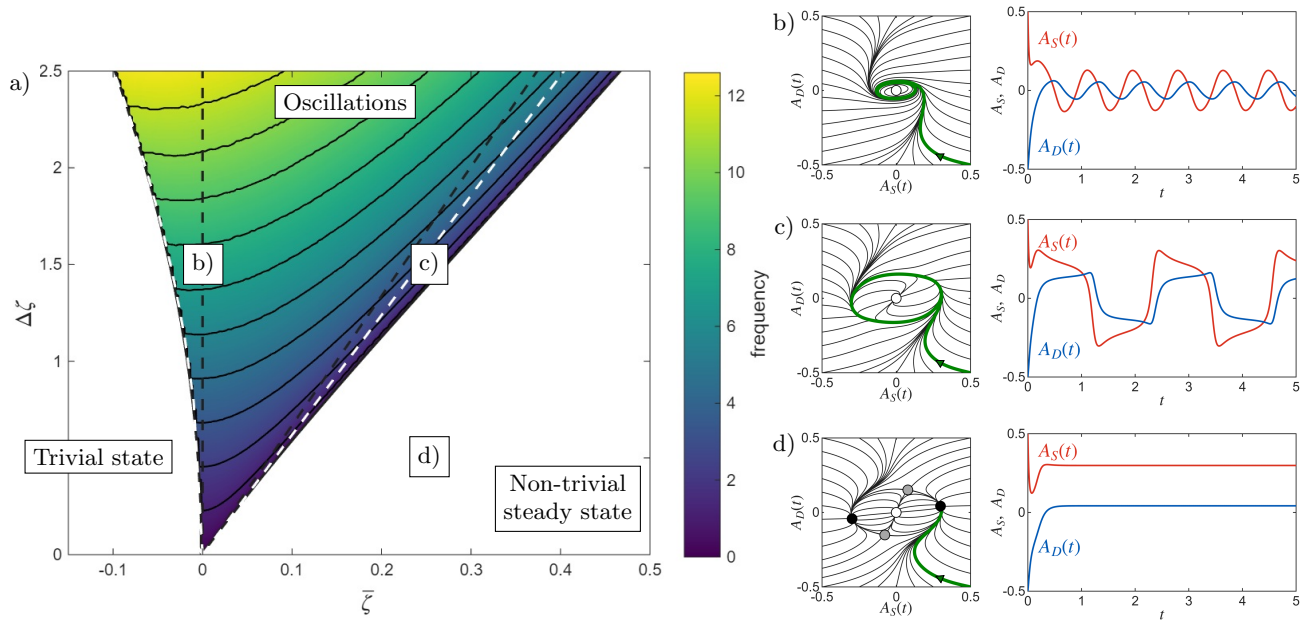

The dynamics in \eqref{eq:DynamSysLin} stem from  \eqref{eq:DirectorDynamics} which, when linearised, contains a term of the form $\partial_t\theta\sim\tilde\zeta\theta$ (see \eqref{eq:R1} in Appendix \ref{AppB}). From this it follows immediately that the $S$ and $D$ modes have their symmetry maintained by even activity perturbations, but are coupled to the mode of opposing symmetry by odd activity perturbations \footnote{At higher orders of perturbation theory, higher-order activity modes would enter. Nonetheless, this symmetry property remains, since the un-linearised term is $\partial_t\theta\sim\zeta\sin(2\theta)$, which has the same symmetry as its linearisation.}. 
The effect of the $\tilde\zeta\theta$ term is illustrated in Figure \ref{fig:GraphicalArgumentForActivityPatterns}. The top row shows the effect of a purely even activity perturbation, namely $\tilde\zeta=\cos\left(4\pi z/d\right)$ so that the only non-zero coefficient in \eqref{eq:LamdaSS} is $a_2$. The signs of $a_2$ in \eqref{eq:LamdaSS} and \eqref{eq:LamdaDD} mean that $A_S$ is damped and $A_D$ is excited, as illustrated in Figure \ref{fig:PhasePortraits}b). The bottom row of Figure \ref{fig:GraphicalArgumentForActivityPatterns} shows the effect of a purely odd activity perturbation, a constant gradient, where only $b_1$ and $b_2$  are non-zero, and since $2b_1>b_2>0$ we see that that non-zero $A_S$ leads to the growth of $A_D$ and non-zero $A_D$ leads to the growth of the $-A_S$. The couplings in this last case are requisite for oscillatory dynamics around the cycle $A_S\to A_D\to -A_S\to -A_D$, as is shown in Figure \ref{fig:PhasePortraits}d), found through numerical simulation of \eqref{eq:DirectorDynamics} for the same activity profile.

\textit{Oscillations due to an activity gradient.} --- 
Of the forms of instability described above, oscillations represent the greatest novelty, and shall be our focus for the remainder of this paper. We do so in the particularly simple context of constant activity gradients, and so consider linear activity profiles of the form
\begin{equation}
    \zeta(z)=\zeta_c\left[1+\bar{\zeta}+\frac{\Delta\zeta}{d}\left(z-\frac{d}{2}\right)\right],
    \label{eq:LinearActivityProfile}
\end{equation}
which incorporate a shift in the average activity alongside the non-zero constant gradient in activity shown in Figure \ref{fig:PhasePortraits}d) and the bottom row of Figure \ref{fig:GraphicalArgumentForActivityPatterns}. Here $\bar{\zeta}$ measures the deviation of the average activity from the critical value and $\Delta\zeta$ captures the difference in activity across the channel. We shall describe the instabilities of a confined active nematic in $(\bar{\zeta},\Delta\zeta)$ space and take $\Delta\zeta>0$, since \eqref{eq:LinearActivityProfile} contains a natural symmetry under $\Delta\zeta\to-\Delta\zeta$ and $z\to d-z$.

Numerical integration of \eqref{eq:DirectorDynamics} identifies three regimes of behaviour, with the boundaries given by the solid lines in Figure \ref{fig:ActivityGradient}a). For sufficiently negative values of $\bar\zeta$, the only stable state is the trivial, undistorted quiescent one; for sufficiently positive values of $\bar\zeta$ the stable state of the system is one of a pair of non-trivial fixed points arranged symmetrically either side of the origin; and for intermediate values of $\bar\zeta$, between the solid curves, the system exhibits sustained oscillations. Representative examples of the oscillations are shown in Figure \ref{fig:ActivityGradient}b) and c), with a non-trivial fixed point illustrated in Figure \ref{fig:ActivityGradient}d). The transition from the trivial state to oscillations occurs upon increasing either the average activity or the difference in activity across the channel, and is mediated by a Hopf bifurcation. When in the region of oscillatory solutions, a further increase of the average activity or a decrease of the difference in activity across the channel can lead to collapse of the limit cycle, with a pair of non-trivial steady states emerging. The classical instability \cite{voituriez2005spontaneous}  corresponds to moving along the horizontal axis of Figure \ref{fig:ActivityGradient}a), with a double pitchfork bifurcation occurring as the system passes through the origin \cite{Walton2020,marzorati2023bifurcation}.

The preceding linear analysis also allows us to divide phase space into three regions, according to whether the trivial state corresponds to a sink or spiral sink; a spiral source; or a source. These three regions are bounded by the black dashed lines in Figure \ref{fig:ActivityGradient}a). However, in comparing the black dashed and solid lines in Figure \ref{fig:ActivityGradient}a) we must be mindful that, while both the transition from the trivial state to a spiral source and the Hopf bifurcation are local processes, the collapse of the limit cycle is a global process, distinct from the spiral source at the origin becoming a source. The black dashed line on the left is therefore the linear approximation to the solid left-hand boundary of the region of oscillations, with asymptotic agreement near the origin of  Figure \ref{fig:ActivityGradient}a). However, the black dashed line on the right is not an approximation of the solid right-hand boundary, and need not be related, even asymptotically. For instance, the formation of additional fixed points within the limit cycle, before the limit cycle collapses, would decouple the behaviour close to the limit cycle and the behaviour around the trivial state, although our numerical calculations show that this is not the case for the parameter values used here. We have also calculated improved approximations to both boundaries of the region of oscillations, which are indicated by the white dashed lines in Figure \ref{fig:ActivityGradient}. The left-hand approximation comes from application of centre manifold theory at quadratic order, while the right-hand approximation arises from the analysis of a resultant, as detailed in Appendix \ref{AppC} and \ref{AppD}, respectively.

Within the region in which sustained oscillations occur, there is considerable variation in the form of the oscillations that the system can display. Broadly speaking, sustained oscillations require both long-time non-zero amplitude and frequency. The Hopf bifurcation curve marks the parameter values at which a long-time non-zero amplitude is achieved, while the boundary of limit cycle collapse marks the parameter values at which a long-time  non-zero frequency is lost. Accordingly, the sustained oscillations that occur close to the Hopf bifurcation, shown in Figure \ref{fig:ActivityGradient}b), have small amplitude. From the Hopf bifurcation, increasing the average activity and approaching the point of limit cycle collapse, the amplitude grows and the frequency falls. Additionally, the oscillations near the Hopf bifurcation are approximately harmonic, while oscillations near to the limit cycle collapse are relaxational \cite{giomi2011excitable,giomi2012banding,giomi2014defect}, exhibiting a fast and slow timescale, with {\it snap-through} behaviour as the system quickly transitions between states on opposite sides of the limit cycle, as seen in Figure \ref{fig:ActivityGradient}c). In this latter situation, although a limit cycle exists, the imprint of the fixed points is discernible in the long periods for which the system remains near certain configurations.

Being close to harmonic, the oscillations that arise near the Hopf bifurcation are amenable to a weakly nonlinear analysis, as detailed in Appendix \ref{AppC}. Taking $\bar{\zeta}\ll1$, the mode amplitudes have the form $A_S=r_S(T)\cos\left(\omega t+\bar{\phi}\right)$, $A_D=r_D(T)\cos\left(\omega t+\bar{\phi}\right)$, where $T=\bar\zeta\omega t$ is a slow timescale and $\bar{\phi}$ is a constant phase shift determined by the initial conditions. The oscillation frequency is given by
\begin{equation}
    \omega=\pi\sqrt{\frac{3}{3+2\Gamma}}\frac{1+\Gamma}{t_e}\Delta\zeta,
    \label{eq:OscillationFrequencyHarmonic}
\end{equation}
while the mode amplitudes are proportional to $\sqrt{\bar\zeta}$, with their asymptotic ratio, as $T\to\infty$, given by $r_S/r_D\to(\sqrt{3}/2)\sqrt{3+2\Gamma}$.

We now consider the behaviour of the relaxation oscillations that occur close to the limit cycle collapse boundary. As detailed in Appendix \ref{AppD}, we numerically investigate the oscillations close to the boundary, finding that the fast timescale is constant as we approach the boundary in parameter space, while the slow timescale, and hence the whole period, is inversely proportional to the square root of the distance from the boundary. The first of these observations can be understood by noting that the fast-timescale jumps in $A_S$ and $A_D$ in Figure \ref{fig:ActivityGradient}c) are offset, such that when one mode snaps, the other is approximately constant. As $A_S$ snaps through zero we have $\dot{A}_S=\Lambda^S_DA_D$, and similarly for $A_D$. Hence, approximating the snap as having constant velocity yields the constant scaling. Although, as mentioned above, the limit cycle collapse is a global phenomena, the scaling of the slow timescale can be shown to emerge from the local behaviour close to the origin of phase space. This approximation of the global behaviour by the local behaviour will be valid close to the classical critical activity (the origin in Figure \ref{fig:ActivityGradient}a)) but is surprisingly robust far from there, as described in Appendix \ref{AppD}.  

More generally, in the region of oscillations, the numerically-calculated oscillation frequency for a given activity profile is indicated by the contours in Figure \ref{fig:ActivityGradient}a). As discussed above, the frequency is maximal near the Hopf bifurcation and decreases to zero at the boundary of limit cycle collapse. Working within the centre manifold analysis calculation, and neglecting the anisotropy of the activity-independent cubic terms, gives an analytically-tractable reduced model which captures the influence of the activity profile on the oscillation frequency (see Appendix \ref{AppD}). This model yields the approximate frequency as
\begin{equation}
    \omega\approx\frac{1}{2}\sqrt{\Delta},
    \label{eq:GeneralFrequencyApproximation}
\end{equation} 
which matches \eqref{eq:OscillationFrequencyHarmonic} when $\bar{\zeta}=0$ and agrees well with the numerically-derived contours shown in Figure \ref{fig:ActivityGradient}a), as demonstrated in the Appendix \ref{AppD}.

Although not the primary focus of this work, we note that just to the right of the boundary of limit cycle collapse in Figure \ref{fig:ActivityGradient}a) the active nematic may display excitable dynamics \cite{giomi2011excitable,giomi2012banding}. This is a consequence of the limit cycle collapse being associated with the emergence of two saddle-sink pairs, shown in gray and black respectively in Figure \ref{fig:ActivityGradient}d). Immediately beyond the limit cycle collapse boundary the saddle and sink are close, such that a perturbation past the saddle will excite the system, causing it to jump from one sink to the opposing one, reversing the director distortion and the active flow.

\textit{Discussion} --- Our investigation of the effect of spatially-heterogeneous activity on the spontaneous flows of active nematics establishes that it can determine the symmetry of the steady state and provides a novel pathway to oscillatory dynamics. While oscillations have previously been found to result through coupling to a viscoelastic medium \cite{mori2023viscoelastic} or a footprint field \cite{bell2025ordering}, our results show that they can arise in an isolated active system. This provides new insight into the types of collective flows that can emerge in biological systems, and the mechanisms behind them, along with new avenues for achieving spatio-temporal control of active matter. In light of this work it is natural to seek an experimental realisation of these oscillations, for example via light-activated active nematics \cite{nakamura2014remote,ross2019controlling,ruijgrok2021optical}. Given the rich dynamics induced by heterogeneous activity, there is also a theoretical challenge to understand more fully the impact this has on active turbulence \cite{assante2023active,partovifard2024controlling,schimming2025turbulence}, as well its role in three-dimensional flows and the chiral structures they exhibit \cite{keogh2022helical,pratley2024three}.

\bibliography{SponFlowHetAcNem.bib}

\begin{widetext}

\newpage

\appendix
\section{Derivation of the Director Dynamics}
\label{AppA}
 Our analysis uses the Ericksen-Leslie equations for an active nematic in terms of its director field $\mathbf{n}$ and fluid flow field $\mathbf{u}$ \cite{de1993physics},
\begin{align}
    \nabla\cdot\mathbf{u}&=0, \label{eq:Incompressable}\\
    0&=-\nabla p+\mu\nabla^2\mathbf{u}+\nabla\cdot\boldsymbol{\sigma},
    \label{eq:Stokes}\\
    \frac{D\mathbf{n}}{Dt}+\mathbf{\Omega}\cdot\mathbf{n}&=\frac{1}{\gamma}\mathbf{h}-\nu\left[\mathbf{D}\cdot\mathbf{n}-(\mathbf{n}\cdot\mathbf{D}\cdot\mathbf{n})\mathbf{n}\right],
    \label{eq:NematoHydro}
\end{align}
where $p$ is the pressure, $\mu$ the Newtonian viscosity, $\boldsymbol{\sigma}$ the total stress, $D/Dt$ the material derivative, $\gamma$ the rotational viscosity, $\mathbf{h}$ the molecular field,  $\nu$ the flow alignment parameter and $D_{ij}=\frac{1}{2}\left(\partial_iu_j+\partial_ju_i\right)$ and $\Omega_{ij}=\frac{1}{2}\left(\partial_iu_j-\partial_ju_i\right)$ are the symmetric and antisymmetric parts of the flow gradient tensor respectively. Here we consider flow-aligning active nematics, for which $\nu<-1$, although we note that non-flow-aligning, or {\it tumbling}, active nematics exhibit interesting behaviour even in the case of homogeneous activity profiles \cite{lavi2025nonlinear}. Within a one elastic constant approximation, the Frank free energy density has the form 
\begin{equation}
    F=\frac{1}{2}\int K\partial_in_j\partial_in_j\text{d}^2\mathbf{r},
\end{equation}
where $K$ is the isotropic elastic constant, and from which the molecular field is found via $h_i=-\delta F/\delta n_i-n_in_j\delta F/\delta n_j$.
The total stress is the sum of elastic and active stresses, given by
\begin{align}
    \sigma_{ij}=&\frac{1}{2}\left(n_ih_j-h_in_j\right)+\frac{\nu}{2}\left(n_ih_j+h_in_j\right)-K\partial_in_k\partial_jn_k-\zeta n_in_j,\label{stress}
\end{align}
where the final term is the active stress and the remaining terms constitute the elastic contribution. The active contribution to the stress tensor is weighted by the activity $\zeta$, which here we consider to be a spatially-varying function, and which is positive for extensile active stresses and negative for contractile ones. 

Although this continuum framework based on the Ericksen-Leslie equations, often using the one elastic constant and flow alignment assumptions, has been extremely successful in demonstrating the fundamental properties of active nematics, there are inevitable limitations. Defects, corresponding to singularities in the director field, are not easily modelled within these equations, and so the commonly observed {\it active turbulence} cannot be reproduced. Effects due to changes in the density or scalar order parameter of the active agents can also not explicitly be included, although we note that a proxy for such changes could be the spatially dependent activity that we consider here. Discussions of alternative models and these additional phenomena can be found, for instance, in the following reviews \cite{doostmohammadi2018active, marchetti2013hydrodynamics, ramaswamy2010mechanics}.

In this work, we assume that the director field remains in a plane, which we define as the $xz$-plane, with translational invariance along the $x$- and $y$-directions, so that $\mathbf{n}=\sin\theta(z)\mathbf{e}_x+\cos\theta(z)\mathbf{e}_z$. Incompressibility then enforces rectilinear flow, such that $\mathbf{u}=u(z)\mathbf{e}_x$. Taking the $z$-component of the Stokes equation \eqref{eq:Stokes} then gives that
\begin{equation}
    p=\sigma_{zz}+p_0,
\end{equation}
where $p_0$ is a constant reference pressure. Then taking the $x$-component of \eqref{eq:Stokes} gives
\begin{equation}
    \partial_z(\mu\partial_zu+\sigma_{xz})=0,
\end{equation}
so that solving for the flow while imposing the no-slip boundary conditions leads to
\begin{equation}
    u(z)=\frac{1}{\mu}\left[\overline{\sigma_{xz}} z-\int_0^z\sigma_{xz}(z')\text{d}z'\right],
    \label{eq:FlowSolution}
\end{equation}
where $\overline{\bullet}$ denotes an average across the channel and from the general expression for the stress tensor given in \eqref{stress} it follows that 
\begin{equation}
    \sigma_{xz}=-\dfrac{K}{2}\left(1-\nu\cos2\theta\right)\partial^2_z\theta-\frac{\zeta}{2}\sin2\theta.
    \label{eq:StressTensorAppenA}
\end{equation}

Projecting \eqref{eq:NematoHydro} along a direction perpendicular to $\mathbf{n}$ yields
\begin{equation}
    \partial_t\theta=\frac{K}{\gamma}\partial^2_z\theta+\frac{1-\nu\cos2\theta}{2}\partial_zu,
    \label{eq:DirectorDynamicsFlow}
\end{equation}
and substituting the flow solution in \eqref{eq:FlowSolution} into \eqref{eq:DirectorDynamicsFlow} produces the director dynamics equation
\begin{equation}
    \partial_t\theta=\frac{K}{\gamma}\partial^2_z\theta+\frac{1-\nu\cos2\theta}{2\mu}\left[\overline{\sigma_{xz}}-\sigma_{xz}(z)\right].
    \label{eq:ThetaDynamicsAppenA}
\end{equation}

A linear stability analysis of \eqref{eq:ThetaDynamicsAppenA} with \eqref{eq:StressTensorAppenA} about the  trivial  state, $\theta\equiv 0$, for a homogeneous activity,  reveals the classic spontaneous flow transition \cite{voituriez2005spontaneous}, in which two modes $\theta_S=\sin\left(2\pi z/d\right)$ and $\theta_D=\frac{1}{2}\left[1-\cos\left(2\pi z/d\right)\right]$ become unstable simultaneously, at a critical activity
\begin{equation}
    \zeta_c=\frac{2\pi^2 K\left[4\mu+\gamma(1-\nu)^2\right]}{d^2\gamma(1-\nu)}=\frac{8\pi^2\mu(1+\Gamma)}{(1-\nu)t_e},
    \label{eq:Zetac}
\end{equation}
where in the second equality we have introduced the elastic timescale $t_e=\gamma d^2/K$ along with the dimensionless constant $\Gamma=\gamma(1-\nu)^2/(4\mu)$. For the values used here, $\nu =-1.1$ and $\gamma/\mu=1$, the bifurcations at $\zeta_c$ are  supercritical pitchforks.

\section{Centre Manifold Analysis}
\label{AppB}
Using centre manifold analysis, we will now determine the amplitude evolution of the $S$ and $D$ modes when the activity is close to the critical homogeneous value, $\zeta_c$. We first take the activity to have the general functional form $\zeta(z)=\zeta_c(1+\varepsilon\Tilde{\zeta}(z))$, restricting $\varepsilon$ to be a small parameter that measures the magnitude of the activity perturbation. Although we later consider the case of a constant gradient in activity across the channel, we first consider this more general form of the spatially-dependent activity profile. The director dynamics \eqref{eq:DirectorDynamics} may then be written as
\begin{equation}
    \partial_t\theta=\mathcal{L}_0\theta+\mathcal{R}_1\theta+O(\varepsilon\theta^3,\theta^5),
    \label{eq:DirectorDynamicsAppenB}
\end{equation}
where $\mathcal{L}_0$ is the linear operator given by 
\begin{equation}
    \mathcal{L}_0\theta=\frac{K}{\gamma}\partial^2_z\theta+\frac{K(1-\nu)^2}{4\mu}\left[\partial^2_z\theta-\overline{\partial^2_z\theta}\right]+\frac{(1-\nu)\zeta_c}{2\mu}\left(\theta-\overline{\theta}\right),
    \label{eq:L0}
\end{equation}
and
\begin{equation}
    \begin{split}
        \mathcal{R}_1\theta&=\frac{K\nu(1-\nu)}{\mu}\left[\theta^2\partial^2_z\theta-\frac{1}{2}\overline{\theta^2\partial^2_z\theta}-\frac{1}{2}\theta^2\overline{\partial^2_z\theta}\right] +\frac{(1-\nu)\zeta_c}{2\mu}\varepsilon\left[\Tilde{\zeta}\theta-\overline{\Tilde{\zeta}\theta}\right]-\frac{1-\nu}{3\mu}\zeta_c\left(\theta^3-\overline{\theta^3}\right)+\frac{\nu}{\mu}\zeta_c\theta^2\left(\theta-\overline{\theta}\right)
    \end{split}
    \label{eq:R1}
\end{equation}
contains the leading order terms coupling the activity perturbation to the director and the cubic terms in the director angle. 
We note that, as follows from the results of \cite{da2012kickback}, the operator $\mathcal{L}_0$ is self-adjoint with respect to the following inner product
\begin{equation}
    \langle f,g\rangle=\int_0^d\left[f(z)+\Gamma\overline{f}\right]g(z)\text{d}z,
    \label{eq:InnerProduct}
\end{equation}
where $\Gamma=\gamma(1-\nu)^2/(4\mu)$. We may now use this inner product to determine the evolution of the amplitudes $A_S(t)$ and $A_D(t)$ of the $S$ and $D$ modes via centre manifold analysis \cite{haragus2010local}. Such an analysis posits that there is a two-dimensional centre manifold upon which the solution to \eqref{eq:DirectorDynamicsAppenB} has the representation
\begin{equation}
    \begin{split}
        \theta=&A_S\theta_S+A_D\theta_D+\varepsilon A_S\Psi_S^{(1)}+\varepsilon A_D\Psi_D^{(1)}+\dots\\
        &+A_S^2\Psi_{SS}^{(0)}+A_SA_D\Psi_{SD}^{(0)}+A_D^2\Psi_{DD}^{(0)}+\varepsilon A_S^2\Psi_{SS}^{(1)}+\dots\\
        &+A_S^3\Psi_{SSS}^{(0)}+\dots,
    \end{split}
    \label{eq:CMTRep}
\end{equation}
with the $\Psi_{\bullet}^{\bullet}$ being functions of $z$ that are perpendicular to the centre manifold and where we are performing an expansion in both the mode amplitudes, $A_S$ and $A_D$, and the small parameter $\varepsilon$. Taking $\theta_0=A_S\theta_S+A_D\theta_D$, it then follows that the dynamics of $A_S$ and $A_D$ are given, up to the same error of the same order as indicated in \eqref{eq:DirectorDynamicsAppenB}, by
\begin{align}
    \dot{A}_S&=\frac{\langle\theta_S,\mathcal{R}_1\theta_0\rangle}{\langle\theta_S,\theta_S\rangle},\label{eq:ASdotR1}\\
    \dot{A}_D&=\frac{\langle\theta_D,\mathcal{R}_1\theta_0\rangle}{\langle\theta_D,\theta_D\rangle}.
    \label{eq:ADdotR1}
\end{align}
Introducing a Fourier series for the activity perturbation
\begin{equation}
    \Tilde{\zeta}=a_0+\sum_{n=1}^{\infty}\left(a_n\cos\left(\frac{2\pi n z}{d}\right)+b_n\sin\left(\frac{2\pi n z}{d}\right)\right),
\end{equation}
the amplitude dynamics are found to be
\begin{align}
    \begin{split}
        \dot{A}_S=&\frac{\varepsilon(1-\nu )\zeta_c}{4\mu }\left[(2a_0-a_2) A_S+\frac{1}{2}(2b_1-b_2) A_D\right]\\
        &-\frac{\zeta_c}{4(1+\Gamma)\mu}\left[\left(1-4\nu+\Gamma(1+2\nu)\right)A_S^3+\frac{5-12\nu+\Gamma(5+2\nu)}{4}A_D^2A_S\right],
        \label{eq:ASCubic}
    \end{split}
    \\
    \begin{split}
        \dot{A}_D=&\frac{\varepsilon(1-\nu)\zeta_c}{2\mu(3+2\Gamma)}
   \left[(a_0-a_1+\frac{1}{2}a_2)A_D-b_2A_S\right]\\
   &-\frac{\zeta_c}{4(1+\Gamma)(3+2\Gamma)\mu}\left[\frac{5-20\nu+\Gamma(5-6\nu)}{4}A_D^3+\left(1-12\nu+\Gamma(1-6\nu)\right)A_S^2A_D\right].
   \label{eq:ADCubic}
    \end{split}
\end{align}
from which we find the linear coefficients given in \eqref{eq:LamdaSS}-\eqref{eq:LamdaDD} after simplification using the expression for the critical activity \eqref{eq:Zetac}. 

\section{Constant Gradient: Hopf Bifurcation Curve}
\label{AppC}
We now concentrate on the particular example of heterogeneous activity, namely a constant gradient in activity about mean activity value,
\begin{equation}
\zeta(z)=\zeta_c\left[1+\bar{\zeta}+\frac{\Delta\zeta}{d}\left(z-\frac{d}{2}\right)\right],
\end{equation}
which, as summarised in the main text, can lead to sustained oscillations of the director when the trivial state is destabilised through a Hopf bifurcation. 

\subsection{Approximation of the Hopf bifurcation curve}
Using the centre manifold analysis described in Appendix \ref{AppB}, we may determine the approximate form of the Hopf bifurcation curve (the left-hand boundary of the region of oscillations in Figure \ref{fig:ActivityGradient}a)) close to the origin of $(\bar\zeta,\Delta\zeta)$ parameter space. To do this we take the constant gradient of the activity profile to parameterise this curve and to be the small parameter $\varepsilon=\Delta\zeta$. We therefore seek the functional form of the Hopf bifurcation curve $\bar\zeta=\bar\zeta_H(\varepsilon)$ so that 
\begin{equation}
    \zeta(z)=\zeta_c\left[1+\bar{\zeta}_H(\varepsilon)+\frac{\varepsilon}{d}\left(z-\frac{d}{2}\right)\right],
    \label{eq:HopfPerturbationGeneral}
\end{equation}
where we expand $\bar{\zeta}_H(\varepsilon)$ as a power series, 
\begin{equation}    \bar{\zeta}_H(\varepsilon)=\varepsilon\bar{\zeta}^{(1)}+\varepsilon^2\bar{\zeta}^{(2)}+\dots.\label{barzetaH}
\end{equation}

Equation \eqref{barzetaH} will be the relationship between an activity gradient and the mean activity which maintain the condition for a Hopf bifurcation, namely $\text{Tr}\mathbf{\Lambda}=0$, where $\mathbf{\Lambda}$ is the matrix defined in \eqref{eq:DynamSysLin}. At linear order in $\varepsilon$ this condition requires that $\bar{\zeta}^{(1)}=0$ (as can be seen by setting $a_0=\bar{\zeta}^{(1)},\,a_1=0=a_2$ in \eqref{eq:LamdaSS} and \eqref{eq:LamdaDD}), and then the lowest order approximation of the curve will be derived from an activity profile of the form
\begin{equation}
    \zeta(z)=\zeta_c\left[1+\frac{\varepsilon}{d}\left(z-\frac{d}{2}\right)+\varepsilon^2\bar{\zeta}^{(2)}\right].
\end{equation}

Following the centre manifold analysis from Appendix \ref{AppB}, we use \eqref{eq:DirectorDynamicsAppenB}, \eqref{eq:L0} and \eqref{eq:R1}, with $\theta$ having the representation given in \eqref{eq:CMTRep}. Working to linear order in the amplitude $A_S$ and $A_D$, we have
\begin{equation}
    \begin{split}
        \partial_t\left(A_S\theta_S+A_D\theta_D \right.& \left.+\varepsilon A_S\Psi_S^{(1)}+\varepsilon A_D\Psi_D^{(1)}+\varepsilon^2 A_S\Psi_S^{(2)}+\varepsilon^2 A_D\Psi_D^{(2)}+\dots \right) = \mathcal{L}_0\left(A_S\theta_S+A_D\theta_D+\varepsilon A_S\Psi_S^{(1)}+\dots\right)\\
        &+\frac{1-\nu}{2\mu}\left[\zeta\left(A_S\theta_S+A_D\theta_D+\varepsilon A_S\Psi_S^{(1)} + \dots\right)-\overline{\zeta\left(A_S\theta_S+A_D\theta_D+\varepsilon A_S\Psi_S^{(1)}+\dots\right)}\right].
        \label{eq:ThetaDotEq}
    \end{split}
\end{equation}
The rates of change of the amplitudes are again given by \eqref{eq:ASdotR1} and \eqref{eq:ADdotR1}, but to track the dependence on $\varepsilon$ explicitly  we expand the coefficients as $\mathbf{\Lambda}=\varepsilon\mathbf{\Lambda}^{(1)}+\varepsilon^2\mathbf{\Lambda}^{(2)}+\dots$, such that
\begin{align}
    \dot{A}_S&=\varepsilon\Lambda_S^{S(1)}A_S+\varepsilon\Lambda_D^{S(1)}A_D+\varepsilon^2\Lambda^{S(2)}_SA_S+\varepsilon^2\Lambda^{S(2)}_DA_D,\label{eq:ASdotNew}\\
    \dot{A}_D&=\varepsilon\Lambda_S^{D(1)}A_S+\varepsilon\Lambda_D^{D(1)}A_D+\varepsilon^2\Lambda_S^{D(2)}A_S+\varepsilon^2\Lambda_D^{D(2)}A_D,
    \label{eq:ADdotNew}
\end{align}
where we have introduced the new notation, $\Lambda_{\bullet}^{\bullet (i)}$, to better keep track of the $\varepsilon^i$ terms. Our ultimate goal is to determine the value of $\bar{\zeta}^{(2)}(\varepsilon)$ which matches the bifurcation curve to quadratic order, i.e. that satisfies the condition
\begin{equation}
    \varepsilon\Lambda_S^{S(1)}+\varepsilon^2\Lambda^{S(2)}_S+\varepsilon\Lambda_D^{D(1)}+\varepsilon^2\Lambda_D^{D(2)}=0.
    \label{eq:HopfCond}
\end{equation}
The $\Lambda_{\bullet}^{\bullet (1)}$ terms follow from \eqref{eq:ASdotR1} and \eqref{eq:ADdotR1}, giving
\begin{align}
    \Lambda_S^{S(1)}&=0,\label{eq:AlphaSS}\\
    \Lambda_D^{S(1)}&=-\frac{3\pi(1+\Gamma)}{2t_e},\label{eq:AlphaSD}\\
    \Lambda_S^{D(1)}&=\frac{2\pi(1+\Gamma)}{(3+2\Gamma)t_e},\label{eq:AlphaDS}\\
    \Lambda_D^{D(1)}&=0.\label{eq:AlphaDD}
\end{align}
Since $\bar{\zeta}^{(2)}$ enters the activity at quadratic order in $\varepsilon$, we must solve \eqref{eq:ThetaDotEq} at second order in $\varepsilon$. This requires us to determine the $\Psi_1^{\bullet}$, which we do by considering \eqref{eq:ThetaDotEq} at first order in $\varepsilon$. Note that at zeroth order in $\varepsilon$ \eqref{eq:ThetaDotEq} reduces to the case of homogeneous activity, leading to the value for the critical activity given in \eqref{eq:Zetac}.

At linear order in $\varepsilon$, \eqref{eq:ThetaDotEq} in conjunction with \eqref{eq:ASdotNew} and \eqref{eq:ADdotNew} give
\begin{equation}
    \begin{split}
        \left(\varepsilon\Lambda_S^{S(1)}A_S+\varepsilon\Lambda_D^{S(1)}A_D\right)\theta_S&+\left(\varepsilon\Lambda_S^{D(1)}A_S+\varepsilon\Lambda_D^{D(1)}A_D\right)\theta_D=\mathcal{L}_0\left(\varepsilon A_S\Psi_S^{(1)}+\varepsilon A_D\Psi_D^{(1)}\right)\\
        &+\frac{1-\nu}{2\mu}\left[\frac{\varepsilon}{d}\left(z-\frac{d}{2}\right)\left(A_S\theta_S+A_D\theta_D\right)-\overline{\frac{\varepsilon}{d}\left(z-\frac{d}{2}\right)\left(A_S\theta_S+A_D\theta_D\right)}\right].
    \end{split}
\end{equation}
As $A_S(t)$ and $A_D(t)$ are distinct functions of time with no a priori relationship, the coefficients of each must balance separately. This yields the following pair of equations
\begin{align}
    \mathcal{L}_0\left(\Psi_S^{(1)}\right)&=\Lambda_S^{S(1)}\theta_S+\Lambda_S^{D(1)}\theta_D-\frac{1-\nu}{2\mu}\left[\frac{\varepsilon}{d}\left(z-\frac{d}{2}\right)\theta_S-\overline{\frac{\varepsilon}{d}\left(z-\frac{d}{2}\right)\theta_S}\right],\\
    \mathcal{L}_0\left(\Psi_D^{(1)}\right)&=\Lambda_D^{S(1)}\theta_S+\Lambda_D^{D(1)}\theta_D-\frac{1-\nu}{2\mu}\left[\frac{\varepsilon}{d}\left(z-\frac{d}{2}\right)\theta_D-\overline{\frac{\varepsilon}{d}\left(z-\frac{d}{2}\right)\theta_D}\right],
\end{align}
which, upon inseting the $\Lambda_{\bullet}^{\bullet (1)}$ values given in \eqref{eq:AlphaSS}-\eqref{eq:AlphaDD}, the form of $\mathcal{L}_0$ given in \eqref{eq:L0} and the expression for the critical activity given in \eqref{eq:Zetac}, may be solved to give
\begin{align}
    \Psi_S^{(1)}&=-\frac{2+\Gamma}{6+4\Gamma}\frac{z}{d}\sin\left(\frac{2\pi z}{d}\right)+\frac{\pi}{2}\cos\left(\frac{2\pi z}{d}\right)\frac{z}{d}\left[\frac{z}{d}-1\right],\\
    \Psi_D^{(1)}&=\frac{1}{2}\frac{z}{d}\left[\cos\left(\frac{2\pi z}{d}\right)-1\right]+\frac{\pi}{4}\sin\left(\frac{2\pi z}{d}\right)\frac{z}{d}\left[\frac{z}{d}-1\right],
\end{align}
which satisfy the boundary conditions appropriate for homeotropic anchoring of the director, $\Psi_{\bullet}^{(1)}(0)=\Psi_{\bullet}^{(1)}(d)=0$.

Turning our attention now to the terms in \eqref{eq:ThetaDotEq} that are second order in $\varepsilon$, we again may consider the coefficients of $A_S$ and $A_D$ seperately, giving
\begin{align}
    \Lambda_S^{S(2)}\theta_S+\Lambda_S^{D(2)}\theta_D+\Lambda_S^{S(1)}\Psi_S^{(1)}+\Lambda_S^{D(1)}\Psi_D^{(1)}&=\mathcal{L}_0\left(\Psi_S^{(2)}\right)+\frac{1-\nu}{2\mu}\left[\frac{1}{d}\left(z-\frac{d}{2}\right)\Psi_S^{(1)}+\bar{\zeta}^{(2)}\theta_S-\overline{\frac{1}{d}\left(z-\frac{d}{2}\right)\Psi_S^{(1)}+\bar{\zeta}^{(2)}\theta_S}\right],\\
    \Lambda_D^{S(2)}\theta_S+\Lambda_D^{D(2)}\theta_D+\Lambda_D^{S(1)}\Psi_S^{(1)}+\Lambda_D^{D(1)}\Psi_D^{(1)}&=\mathcal{L}_0\left(\Psi_D^{(2)}\right)+\frac{1-\nu}{2\mu}\left[\frac{1}{d}\left(z-\frac{d}{2}\right)\Psi_D^{(1)}+\bar{\zeta}^{(2)}\theta_D-\overline{\frac{1}{d}\left(z-\frac{d}{2}\right)\Psi_D^{(1)}+\bar{\zeta}^{(2)}\theta_D}\right].
\end{align}
The four $\Lambda_{\bullet}^{\bullet (2)}$ may be determined by taking the inner product of these two equations with $\theta_S$ and $\theta_D$, with the inner product defined in \eqref{eq:InnerProduct}. In each case, the $\Psi_{\bullet}^{(2)}$ do not need to be determined, since $\langle\theta_S,\mathcal{L}_0(\Psi_{\bullet}^{(2)})\rangle=\langle\Psi_{\bullet}^{(2)},\mathcal{L}_0(\theta_S)\rangle=0,\,\langle\theta_D,\mathcal{L}_0(\Psi_{\bullet}^{(2)})\rangle=\langle\Psi_{\bullet}^{(2)},\mathcal{L}_0(\theta_D)\rangle=0$. In this way we find
\begin{align}
    \Lambda_S^{S(2)}&=\frac{8\pi^2(1+\Gamma)}{t_e}\left(\frac{3+5\Gamma}{65\pi^2(3+2\Gamma)}-\frac{1}{96}+\frac{1}{2}\bar{\zeta}^{(2)}\right),\\
    \Lambda_S^{D(2)}&=\frac{\pi(1+\Gamma)(4+3\Gamma)}{2(3+2\Gamma)^2t_e},\\
    \Lambda_D^{S(2)}&=\frac{3\pi(1+\Gamma)(4+3\Gamma)}{8(3+2\Gamma)t_e},\\
    \Lambda_D^{D(2)}&=\frac{4\pi^2(1+\Gamma)}{(3+2\Gamma)t_e}\left(\frac{129+143\Gamma+36\Gamma^2}{32\pi^2(3+2\Gamma)}-\frac{1}{48}+\bar{\zeta}^{(2)}\right).
\end{align}

Returning to our condition for the Hopf bifurcation in \eqref{eq:HopfCond}, we now have
\begin{equation}
    \bar{\zeta}^{(2)}=\frac{1}{48}-\frac{69+82\Gamma+23\Gamma^2}{32\pi^2(2+\Gamma)(3+2\Gamma)}.
\end{equation}
Thus we have found an approximate expression for the Hopf bifurcation curve, 
\begin{equation}
    \bar{\zeta}_H\approx \varepsilon\bar{\zeta}^{(1)}+\varepsilon^2\bar{\zeta}^{(2)}=(\Delta\zeta)^2\left(\frac{1}{48}-\frac{69+82\Gamma+23\Gamma^2}{32\pi^2(2+\Gamma)(3+2\Gamma)}\right).
\end{equation}
which we plot in Figure \ref{fig:ActivityGradient}a) as a white dashed curve. Since $\Gamma>0$, the value of  $\bar{\zeta}^{(2)}$ is always negative, corresponding to the leftwards curve of the boundary in Figure \ref{fig:ActivityGradient}, and meaning that constant gradients reduce the total activity required for the onset of instability. This effect is maximised for $\Gamma=\sqrt{3}$.

\subsection{Weakly nonlinear analysis close to the Hopf bifurcation}
\label{subsec:WNL}
In order to investigate the oscillatory dynamics of the director close to the Hopf bifurcation curve, we now consider a perturbation of mean activity away from the value at the Hopf bifurcation, so that $\bar\zeta=\delta+\bar\zeta^{(2)}\varepsilon^2$ ($\delta\gg\varepsilon^2$) with $\Delta\zeta=\varepsilon$. The leading order terms in the activity profile are therefore,
\begin{equation}
    \zeta(z)=\zeta_c\left[1+\delta+\frac{\varepsilon}{d}\left(z-\frac{d}{2}\right)\right].
\end{equation}
Assuming $\delta$ remains small relative to $\varepsilon$, so that $\delta\ll\varepsilon$, the equations for the evolution of $A_S$ and $A_D$ contain nonlinear terms when $A_S,\,A_D$ are of order $\sqrt{\delta}$ so that the dynamics are given by
\begin{align}
    \dot{A}_S&=\Lambda^S_SA_S+\Lambda^S_DA_D+\Lambda^S_{SSS}A_S^3+\Lambda^S_{DDS}A_D^2A_S
    \label{eq:ASLambdas},\\
    \dot{A}_D&=\Lambda^D_SA_S+\Lambda^D_DA_D+\Lambda^D_{DDD}A_D^3+\Lambda^D_{SSD}A_S^2A_D,
    \label{eq:ADLambdas}
\end{align}
with linear coefficients
\begin{equation}
    \begin{split}
        \Lambda_{S}^{S}&=\frac{4\pi^2(1+\Gamma)}{t_e}\delta,\\
        \Lambda_{S}^{D}&=-\frac{3\pi(1+\Gamma)}{2t_e}\varepsilon,\\
        \Lambda_{D}^{S}&=\frac{2\pi(1+\Gamma)}{(3+2\Gamma)t_e}\varepsilon,\\
        \Lambda_{D}^{D}&=\frac{4\pi^2(1+\Gamma)}{(3+2\Gamma)t_e}\delta,
    \end{split}
    \label{eq:LinearCoeffsGradient}
\end{equation}
and cubic coefficients as provided in \eqref{eq:ASCubic} and \eqref{eq:ADCubic}. 

Using the scaled variables $\mathfrak{A}_S=A_S/\sqrt{\delta},\  \mathfrak{A}_D=A_D/\sqrt{\delta}$, and differentiating \eqref{eq:ASLambdas} and \eqref{eq:ADLambdas}, we obtain the second order differential equations
\begin{align}
    \ddot{\mathfrak{A}}_S&=\Lambda^S_D\dot{\mathfrak{A}}_D+\delta\left[2\Lambda^S_{DDS}\mathfrak{A}_S\mathfrak{A}_D\dot{\mathfrak{A}}_D+\left(\Lambda^S_S+\Lambda^S_{DDS}\mathfrak{A}_D^2+3\Lambda^S_{SSS}\mathfrak{A}_S^2\right)\dot{\mathfrak{A}}_S\right],\label{eq:gothAS}
    \\
    \ddot{\mathfrak{A}}_D&=\Lambda^D_S\dot{\mathfrak{A}}_S+\delta\left[2\Lambda^D_{SSD}\mathfrak{A}_D\mathfrak{A}_S\dot{\mathfrak{A}}_S+\left(\Lambda^D_D+\Lambda^D_{SSD}\mathfrak{A}_S^2+3\Lambda^D_{DDD}\mathfrak{A}_D^2\right)\dot{\mathfrak{A}}_D\right].\label{eq:gothAD}
\end{align}

To leading order, \eqref{eq:gothAS} and \eqref{eq:gothAD} correspond to harmonic oscillators, $\dot{\mathfrak{A}}_S=\Lambda^S_D\mathfrak{A}_D,\,\dot{\mathfrak{A}}_D=\Lambda^D_S\mathfrak{A}_S$, with frequency 
\begin{equation}
    \omega=\sqrt{-\Lambda^S_D\Lambda^D_S}=\pi\sqrt{\frac{3}{3+2\Gamma}}\frac{1+\Gamma}{t_e}\varepsilon,
\end{equation}
and so we now introduce a dimensionless scaled time $\tau=\omega t$.

We may use \eqref{eq:ASLambdas} and \eqref{eq:ADLambdas} to decouple our system of equations, so that to leading order in $\delta$ we have
\begin{align}
    \dfrac{{\rm d}^2 \mathfrak{A}_S}{{\rm d}\tau^2}+\mathfrak{A}_S-\frac{\delta}{\omega}\dfrac{{\rm d} \mathfrak{A}_S}{{\rm d}\tau}\left[\Lambda^S_S+\Lambda^D_D+\left(\Lambda^D_{SSD}+2\frac{\Lambda^D_S}{\Lambda^S_D}\Lambda^S_{DDS}+3\Lambda^S_{SSS}\right)\mathfrak{A}_S^2+\left(\frac{\omega}{\Lambda^S_D}\right)^2\left(\Lambda^D_{DDD}+\Lambda^S_{DDS}\right)\dfrac{{\rm d} \mathfrak{A}_S}{{\rm d}\tau}\right]&=0,
    \label{eq:ASWNL}\\
    \dfrac{{\rm d}^2 \mathfrak{A}_D}{{\rm d}\tau^2}+\mathfrak{A}_D-\frac{\delta}{\omega}\dfrac{{\rm d} \mathfrak{A}_D}{{\rm d}\tau}\left[\Lambda^S_S+\Lambda^D_D+\left(\Lambda^S_{DDS}+2\frac{\Lambda^S_D}{\Lambda^D_S}\Lambda^D_{SSD}+3\Lambda^D_{DDD}\right)\mathfrak{A}_D^2+\left(\frac{\omega}{\Lambda^D_S}\right)^2\left(\Lambda^S_{SSS}+\Lambda^D_{SSD}\right)\dfrac{{\rm d} \mathfrak{A}_D}{{\rm d}\tau}\right]&=0.
    \label{eq:ADWNL}
\end{align}

Both of these equations are now in the form of a generic weakly nonlinear oscillator of the form $x''+x+\delta h(x,x')=0$, the solutions of which can be written as
\begin{align}
    \mathfrak{A}_S&=r_S(T)\cos(\tau+\bar{\phi}(T)),\\
    \mathfrak{A}_D&=r_D(T)\sin(\tau+\bar{\phi}(T)).
\end{align}
where $T=\delta\tau$ is a slow timescale and the slow dynamics of the amplitude, $r_{\bullet}(T)$, and  phase, $\bar\phi(T)$, are then given by the averaged equations \cite{strogatz2024nonlinear}
\begin{align}
    \frac{\text{d}r_{\bullet}}{\text{d}T}&=\frac{1}{2\pi}\int_0^{2\pi}h_{\bullet}\left(\mathfrak{A}_{\bullet},\dfrac{{\rm d} \mathfrak{A}_{\bullet}}{{\rm d}\tau}\right)\sin(\tau+\bar{\phi}(T))\,\text{d}\tau,\label{rdynamics}\\
    r_{\bullet}\frac{\text{d}\bar{\phi}}{\text{d}T}&=\frac{1}{2\pi}\int_0^{2\pi}h_{\bullet}\left(\mathfrak{A}_{\bullet},\dfrac{{\rm d} \mathfrak{A}_{\bullet}}{{\rm d}\tau}\right)\cos(\tau+\bar{\phi}(T))\,\text{d}\tau,\label{phidynamics}
\end{align}
where the $h_{\bullet}$ are the order $\delta$ terms in \eqref{eq:ASWNL} and \eqref{eq:ADWNL}.

We set initial conditions $A_S(0)=S_0,\,A_D(0)=D_0$, so that $\mathfrak{A}_S(0)=S_0/\sqrt{\delta}$ and $\mathfrak{A}_D=D_0/\sqrt{\delta}$, together with those implied by the scaled version of \eqref{eq:ASLambdas} and \eqref{eq:ADLambdas}, namely
\begin{align}
    \dfrac{{\rm d} \mathfrak{A}_S}{{\rm d}\tau}(0)&=\frac{1}{\sqrt{\delta}}\left(\Lambda^S_DD_0+\Lambda^S_{SSS}S_0^3+\Lambda^S_{DDS}S_0D_0^2+\varepsilon\Lambda^S_SS_0\right),
    \label{eq:ASdotInitial}\\
   \dfrac{{\rm d} \mathfrak{A}_D}{{\rm d}\tau}(0)&=\frac{1}{\sqrt{\delta}}\left(\Lambda^D_SS_0+\Lambda^D_{DDD}D_0^3+\Lambda^D_{SSD}D_0S_0^2+\varepsilon\Lambda^D_DD_0\right).
    \label{eq:ADdotInitial}
\end{align}

From \eqref{phidynamics} we find that the phase shift $\bar{\phi}$ is independent of $T$, while using \eqref{rdynamics} with \eqref{eq:ASWNL} gives that the $S$ mode amplitude evolves according to
\begin{equation}
    r_S(T)=2S_0\sqrt{\frac{\Lambda^S_D(\Lambda^S_S+\Lambda^D_D)}{S_0^2\left[\Lambda^D_S\left(\Lambda^S_{DDS}+3\Lambda^D_{DDD}\right)-\Lambda^S_D\left(\Lambda^D_{SSD}+3\Lambda^S_{SSS}\right)\right]\left(1-\text{e}^{-{\rm Tr}{\mathbf\Lambda}  T/\omega}\right)+4\Lambda^S_D\left(\Lambda^S_S+\Lambda^D_D\right)\text{e}^{-{\rm Tr}{\mathbf\Lambda} T/\omega}}},
\end{equation}
such that the asymptotic amplitude, as $T\rightarrow\infty$, is
\begin{equation}
    r_S^{\infty}=2\sqrt{\frac{\Lambda^S_D\left(\Lambda^S_S+\Lambda^D_D\right)}{\Lambda^D_S\left(\Lambda^S_{DDS}+3\Lambda^D_{DDD}\right)-\Lambda^S_D\left(\Lambda^D_{SSD}+3\Lambda^S_{SSS}\right)}}.
\end{equation}
The corresponding expressions for $r_D$ are obtained by exchanging $S$ and $D$, and substituting the explicit expressions for the $\Lambda$ coefficients yields the following ratio of the asymptotic mode amplitudes
\begin{equation}
    \frac{r_S^{\infty}}{r_D^{\infty}}=\frac{\sqrt{3}}{2}\sqrt{3+2\Gamma}.
\end{equation}
The phase shift $\bar{\phi}$ is determined from the initial conditions via projection onto the limit cycle, such that
\begin{equation}
    \bar{\phi}=\tan^{-1}\left(\frac{D_0/r_D^{\infty}}{S_0/r_S^{\infty}}\right).
\end{equation}

The weakly nonlinear solutions for $A_S$ and $A_D$ agree extremely well with the solutions of the cubic evolution equations, \eqref{eq:ASLambdas} and \eqref{eq:ADLambdas}, and, as can be seen in Figure \ref{fig:CubicWNLComparisonModes}, this agreement is retained even for larger values of the constant gradient $\Delta\zeta$ and for relatively large initial values of the amplitudes. Similarly good agreement can found with the full numerical solutions of \eqref{eq:DirectorDynamics}, such as the solutions shown in Figure \ref{fig:ActivityGradient}b). 

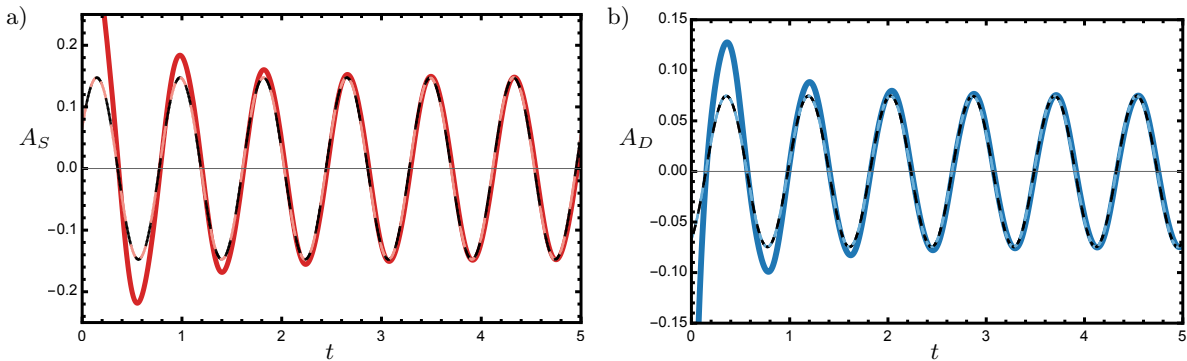
\begin{figure*}
    \centering
    \begin{tikzpicture}[scale=1.6,>=stealth]
        \node[anchor=south west,inner sep=0] at (0,0)
{\includegraphics[width=0.4\linewidth, trim = 0 0 0 0, clip, angle = 0, origin = c]{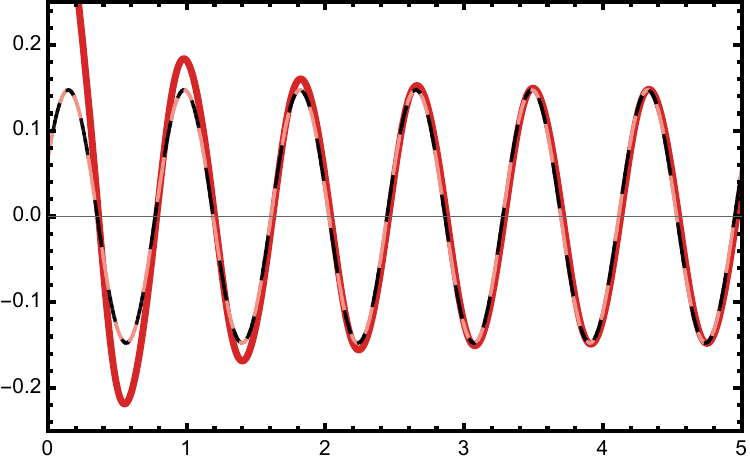}};
        \node[anchor=south west,inner sep=0] at (5,0)
{\includegraphics[width=0.4\linewidth, trim = 0 0 0 0, clip, angle = 0, origin = c]{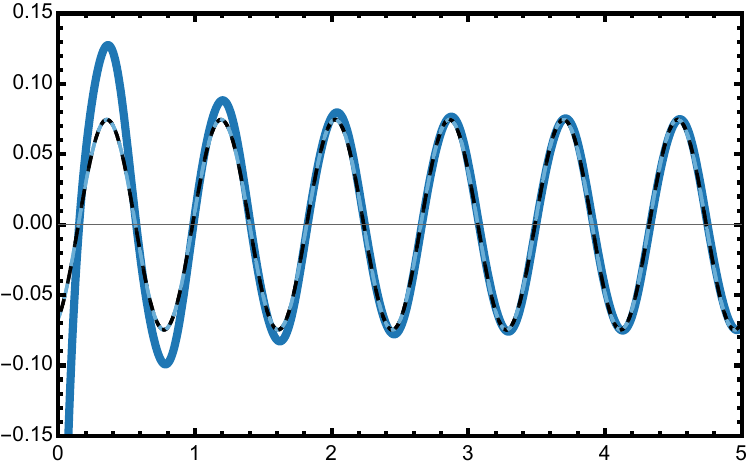}};

        \node[scale=1] at (-0.1,1.75) {$A_S$};
        \node[scale=1] at (2.35,-0.025) {$t$};
        \node[scale=1] at (4.9,1.75) {$A_D$};
        \node[scale=1] at (7.375,-0.025) {$t$};
        \node[scale=1] at (-0.25,2.75) {a)};
        \node[scale=1] at (4.75,2.75) {b)};
    \end{tikzpicture}
    \caption[The \textcolor{red}{unstable mode associated with the} spontaneous flow transition for a homogeneous active nematic film.]{The comparison of the numerical solution of the cubic growth rate equations, \eqref{eq:ASLambdas} and \eqref{eq:ADLambdas}, with the asymptotic form of the solutions derived from the weakly nonlinear analysis. The numerical solution is shown as thick and solid lines, with the weakly nonlinear solution as dashed lines, for a) the $A_S$ mode and b) the $A_D$ mode, with $\Delta\zeta=1.5$, $\bar{\zeta}=0.01$ and the initial conditions $A_S(0)=0.5$, $A_D(0)=-0.5$.
    }
    \label{fig:CubicWNLComparisonModes}
\end{figure*}

\section{Constant Gradient: Limit Cycle Collapse}
\label{AppD}
When in the region of oscillatory solutions shown in Figure \ref{fig:ActivityGradient}a), a further increase of the average activity or a decrease of the difference in activity across the channel can lead to collapse of the limit cycle, with a pair of non-trivial steady states emerging. On the boundary between the region of oscillatory solutions and the region of non-trivial steady states the limit cycle becomes a heteroclinic cycle.

\subsection{Approximating the limit cycle collapse boundary}
We now approximate, using the cubic approximation within the centre manifold analysis given by \eqref{eq:ASCubic} and \eqref{eq:ADCubic}, the curve in $(\bar{\zeta},\Delta\zeta)$ space along which the limit cycle collapse occurs. The key to this approximation is the observation that the transition between oscillations and a non-trivial steady state is marked by the coincidence of two stationary points, a sink and a saddle, in $(A_S,A_D)$ space, examples of which are shown in Figure \ref{fig:PhasePortraits}a) and Figure \ref{fig:ActivityGradient}d). Any such stationary point corresponds to both $\dot{A}_S=0$ and $\dot{A}_D=0$. With $\dot{A}_S$ and $\dot{A}_D$ given by \eqref{eq:ASCubic} and \eqref{eq:ADCubic}, we have two coupled cubic polynomials equations, which are simultaneously satisfied when the resultant of these equations vanishes. We can take this resultant with respect to either $A_S$ or $A_D$, and here choose $A_S$ without loss of generality, so that 
\begin{equation}
    \text{Res}_{A_S}\left(\dot{A}_S,\dot{A}_D\right)=0
    \label{eq:Resultant1}
\end{equation}
provides an equation in $A_D$, the solutions of which correspond to stationary points in $(A_S,A_D)$ space.

To have coincident stationary points requires a repeated root of \eqref{eq:Resultant1}. This corresponds to not only the resultant vanishing, but also its derivative, that is $\partial_{A_S}\left[\text{Res}_{A_S}\left(\dot{A}_S,\dot{A}_D\right)\right]=0$. This once again leaves us with two conditions to be simultaneously satisfied, which may be achieved when the appropriate resultant is zero. The condition for coincident stationary points is therefore
\begin{equation}
    \text{Res}_{A_S}\left(\text{Res}_{A_S}\left(\dot{A}_S,\dot{A}_D\right),\partial_{A_S}\left[\text{Res}_{A_S}\left(\dot{A}_S,\dot{A}_D\right)\right]\right)=0. \label{eq:Resultant2}
\end{equation}
With the Fourier coefficients for the activity profile given by \eqref{eq:LinearActivityProfile} substituted in \eqref{eq:ASCubic} and \eqref{eq:ADCubic}, this condition produces a sixth-order polynomial in $(\Delta\zeta/\bar{\zeta})^2$, which may readily be obtained using a mathematical algebra package, but is a lengthy expression in terms of the various parameters and so we do not show here for brevity. The fact that \eqref{eq:Resultant2} is a polynomial equation for $(\Delta\zeta/\bar{\zeta})^2$ means that the approximation produces straight lines in $(\bar{\zeta},\Delta\zeta)$ space, as shown by the right-hand white dashed line in Figure \ref{fig:ActivityGradient}a). It is apparent from Figure \ref{fig:ActivityGradient}a) that this approximation of the limit cycle collapse boundary is not as satisfactory for larger values of $\bar{\zeta}$ or $\Delta\zeta$ than that of the Hopf bifurcation boundary in Appendix \ref{AppC}. This might be expected since for this resultant calculation we are approximating a non-trivial ground state, whereas in the case of the Hopf bifurcation we are approximating about the trivial state, $\theta\equiv 0$ for any value of $\Delta\zeta$. 

\subsection{Relaxation oscillations close to the limit cycle collapse boundary}
We now turn our attention to the relaxation oscillations that arise when the system is close to the limit cycle collapse boundary. Specifically, we wish to determine how the slow and fast timescales apparent in Figure \ref{fig:ActivityGradient}c) scale with distance to this boundary, which we again denote as $\delta$ as in the case of the Hopf bifurcation dynamics.
However, a rigorous analysis is not possible in this case since, as mentioned above, the ground state is not known analytically. We therefore begin by probing these scalings numerically, examining the oscillations at points in $(\bar{\zeta},\Delta\zeta)$ space along a line perpendicular to the limit cycle collapse boundary. Figure \ref{fig:RelaxationOscillationsAnalysis}a) shows the oscillations at twenty such points, with the distance from the limit cycle collapse boundary varying linearly from $\delta\sim 0.02$ (cyan) to $\delta\sim 0.0005$ (magenta). The oscillations have been shifted so that the point where $A_S=0$ is coincident, allowing timescales to be compared directly. These results reveals that, while the slow timescale increases as the boundary is approached, the fast timescale remains approximately constant. Figure \ref{fig:RelaxationOscillationsAnalysis}b) shows the period, which we denote by $P$, against distance from the boundary $\delta$, on a log-log plot, revealing that $P$, and hence the slow timescale, scale as $1/\sqrt{\delta}$.

The constancy of the fast timescale may be explained by a simple argument based on two observations. Firstly, from Figure \ref{fig:RelaxationOscillationsAnalysis}a) it seems reasonable to approximate the {\it snap-through} behaviour, where $A_S$ quickly transitions between positive and negative values, as straight lines so that $A_S$ changes with constant speed. Secondly, from Figure \ref{fig:ActivityGradient}c) we see that the snap-through in the two modes are offset, such that during a change in one amplitude, the other is approximately constant. Therefore, we take the fast timescale to be determined by the velocity when the snap-through crosses zero. From \eqref{eq:ASCubic} this gives
\begin{equation}
    \dot{A}_S=\Lambda^S_DA_D,
\end{equation}
with the similar equation for $A_D$. While it is true that $\Lambda^S_D$ and the relevant magnitude of $A_D$ depend on activity, and hence $\delta$, these corrections are at first order in $\delta$. Therefore, the snap velocity is a constant up to terms that are $O(\delta)$ and the invariance of the fast timescale is attained.

\begin{figure*}[ht]
    \centering
    \begin{tikzpicture}[scale=1.6,>=stealth]
        \node[anchor=south west,inner sep=0] at (0,0)
{\includegraphics[width=0.425\linewidth, trim = 0 0 0 0, clip, angle = 0, origin = c]{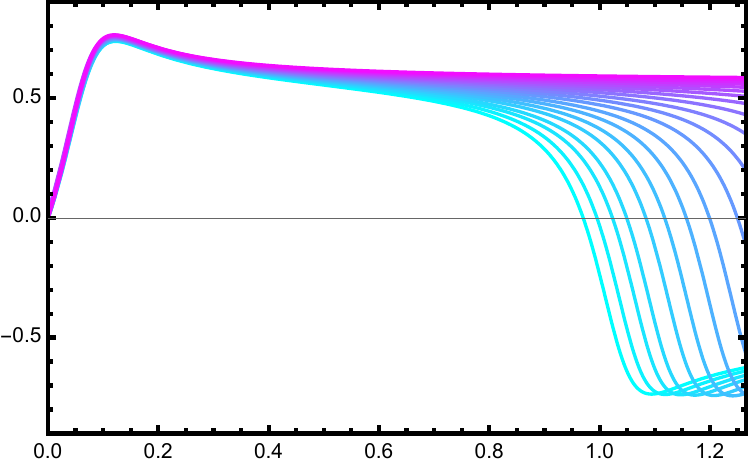}};
        \node[anchor=south west,inner sep=0] at (6,0)
{\includegraphics[width=0.41\linewidth, trim = 0 0 0 0, clip, angle = 0, origin = c]{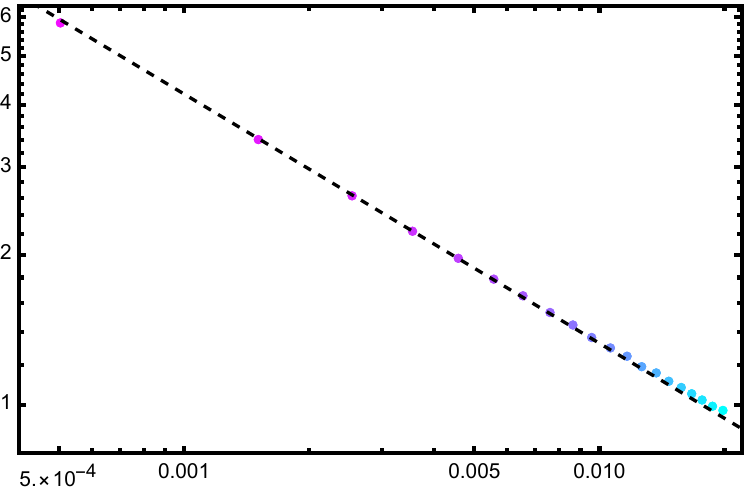}};

        \node[scale=1] at (-0.1,1.75) {$A_S$};
        \node[scale=1] at (2.6,-0.125) {$t$};
        \node[scale=1] at (5.75,1.75) {$P$};
        \node[scale=1] at (8.25,-0.125) {$\delta$};
        \node[scale=1] at (-0.15,2.9) {a)};
        \node[scale=1] at (5.6,2.9) {b)};
        \node[scale=1] at (7.65,2.5) {$\delta^{-1/2}$};
        
    \end{tikzpicture}
    \caption[The behaviour of the relaxation oscillations as a function of distance from the limit cycle collapse boundary.]{The behaviour of the relaxation oscillations as a function of distance from the limit cycle collapse boundary. a) Aligned oscillations for points in $(\bar{\zeta},\Delta\zeta)$ space whose distance from the limit collapse boundary varies linearly from $\sim0.02$ (cyan) to $\sim0.0005$ (magenta). The short timescale is invariant, while the long timescale grows with decreasing distance. b) The period of the same oscillations as a function of distance from the limit cycle collapse boundary, $\delta$, showing that the period, and hence long timescale, grows as $\delta^{-1/2}$. In both panels time is non-dimensionalised by the elastic timescale $t_e=\gamma d^2/K$.
    }
    \label{fig:RelaxationOscillationsAnalysis}
\end{figure*}
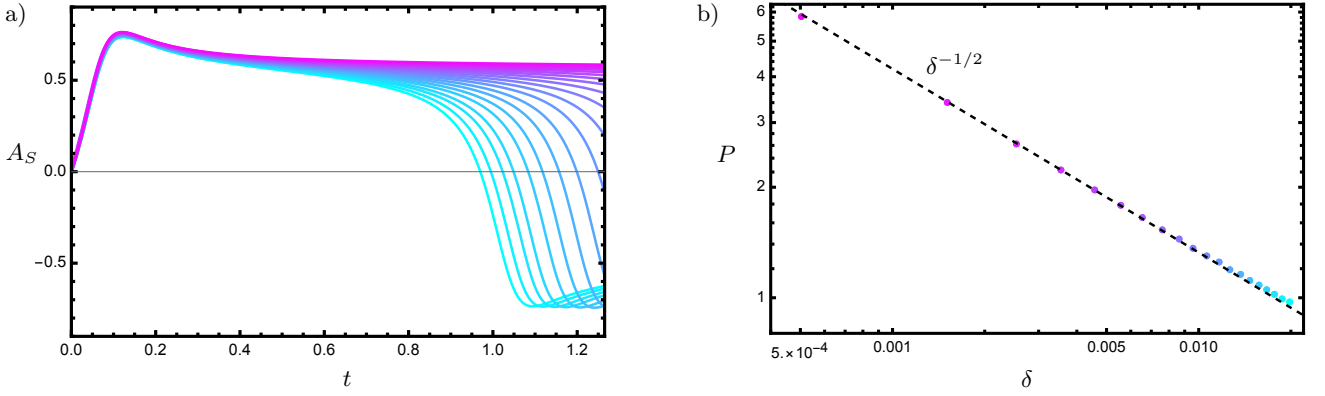

We now consider how the slow timescale scales with $\delta$. We note that all the cubic terms in \eqref{eq:ASCubic} and \eqref{eq:ADCubic} are independent of activity and are not needed to recover the scaling, and so to simplify the analysis we take a reduced isotropic form, such that the cubic terms only affect the radial dynamics. We now decompose the linear terms in the following way
\begin{equation}
    \mathbf{\Lambda}=\alpha_{\radialarrows}\begin{pmatrix}
        1 & 0\\
        0 & 1
    \end{pmatrix}
    +\alpha_{\hyponearrows}\begin{pmatrix}
        1 & 0\\
        0 & -1
    \end{pmatrix}
    +\alpha_{\hyptwoarrows}\begin{pmatrix}
        0 & 1\\
        1 & 0
    \end{pmatrix}
    +\alpha_{\circlearrowleft}\begin{pmatrix}
        0 & -1\\
        1 & 0
    \end{pmatrix},
    \label{eq:LambdaDecomposition}
\end{equation}
so that ${\rm Tr}\Lambda=2\alpha_{\radialarrows}$ and ${\rm det}\Lambda=\alpha_{\radialarrows}^2-\alpha_{\hyponearrows}^2-\alpha_{\hyptwoarrows}^2+\alpha_{\circlearrowleft}^2$.
Using polar coordinates in the $(A_s,\,A_D)$ plane, so that $A_S=r\cos\phi$ and $A_D=r\sin\phi$, the polar angle dynamics of the system is found to be of the form
\begin{equation}
    \dot{\phi}=\alpha_{\circlearrowleft}+\sqrt{\alpha_{\hyponearrows}^2+\alpha_{\hyptwoarrows}^2}\cos2(\phi+\phi_0),
    \label{eq:ToyModelPhiDynamics}
\end{equation}
where $\phi_0=\frac{1}{2}\tan^{-1}(\alpha_{\hyponearrows}/\alpha_{\hyptwoarrows})$. The period is then given by
\begin{equation}
    P=\int_0^{2\pi}\frac{\text{d}\phi}{\dot{\phi}}=\frac{2\pi}{\sqrt{\alpha_{\circlearrowleft}^2-\alpha_{\hyponearrows}^2-\alpha_{\hyptwoarrows}^2}}.
    \label{eq:OscillationPeriod}
\end{equation}
Within this simplified, isotropic cubic theory the limit cycle collapse occurs as $P\rightarrow\infty$, so when $\alpha_{\circlearrowleft}^2\rightarrow\alpha_{\hyponearrows}^2-\alpha_{\hyptwoarrows}^2$, with the emergent steady states of \eqref{eq:ToyModelPhiDynamics} at $\phi=\pm\pi/2-\phi_0$ which replace the limit cycle. At a distance $\delta$ of this limit cycle collapse boundary we set $\alpha_{\circlearrowleft}-\sqrt{\alpha_{\hyponearrows}^2+\alpha_{\hyptwoarrows}^2}=\delta$ in \eqref{eq:OscillationPeriod} and find
\begin{equation}
    P=\pi\sqrt{\frac{2}{\alpha_{\circlearrowleft}\delta}}+O(\sqrt{\delta}),\label{Peq}
\end{equation}
recovering the scaling illustrated in Figure \ref{fig:RelaxationOscillationsAnalysis}b). We note that this is the expected generic scaling for a non-uniform oscillator \cite{strogatz2024nonlinear}.

In fact, the expression for the period of oscillations \eqref{eq:OscillationPeriod} compres well to the full numerical solution far from the limit cycle collapse boundary. Since $\Delta=4\,\text{det}\boldsymbol{\Lambda}-\left(\text{Tr}\boldsymbol{\Lambda}\right)^2$, and given the expressions for $\text{Tr}\boldsymbol{\Lambda}$ and $\text{det}\boldsymbol{\Lambda}$ after \eqref{eq:LambdaDecomposition}, then \eqref{eq:OscillationPeriod} gives the frequency $\omega=2\pi/P$ as 
\begin{equation}
    \omega=\frac{1}{2}\sqrt{\Delta}.\label{eqomega}
\end{equation}
This frequency is plotted, after non-dimensionalising by the elastic timescale $\gamma d^2/K$, in Figure \ref{fig:FrequencyApproximation}a). Since this approximation derives from the linear theory, we have restricted the plot to the region in $(\bar\zeta,\,\Delta\zeta)$ parameter plane for which the linear dynamics produce an unstable spiral node. The numerically obtained frequency derived from solutions of the full nonlinear equation \eqref{eq:DirectorDynamics} is shown in Figure \ref{fig:FrequencyApproximation}b), and we find good agreement between the approximate expression in \eqref{eqomega} and the numerical results in both magnitude of the frequency and the hyperbolic shape of the frequency contours.

\begin{figure*}
    \centering
    \begin{tikzpicture}[scale=1.6,>=stealth]
        \node[anchor=south west,inner sep=0] at (0,0)
{\includegraphics[width=0.425\linewidth, trim = 120 290 120 300, clip, angle = 0, origin = c]{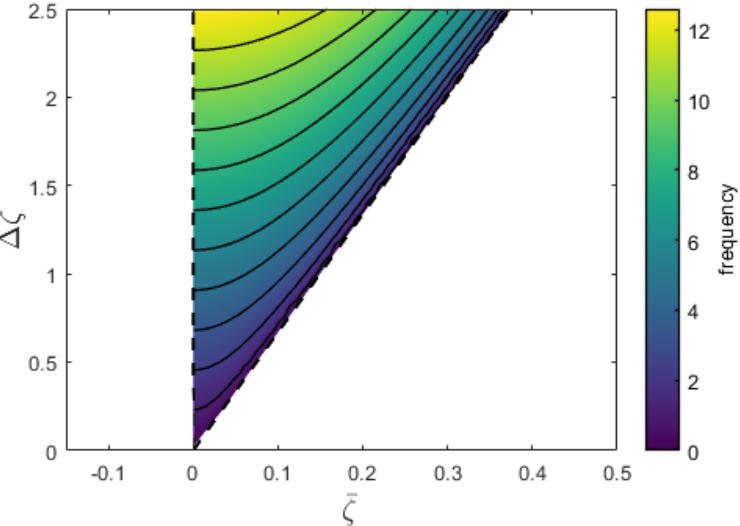}};
        \node[anchor=south west,inner sep=0] at (5.75,0)
{\includegraphics[width=0.425\linewidth, trim = 120 290 120 300, clip, angle = 0, origin = c]{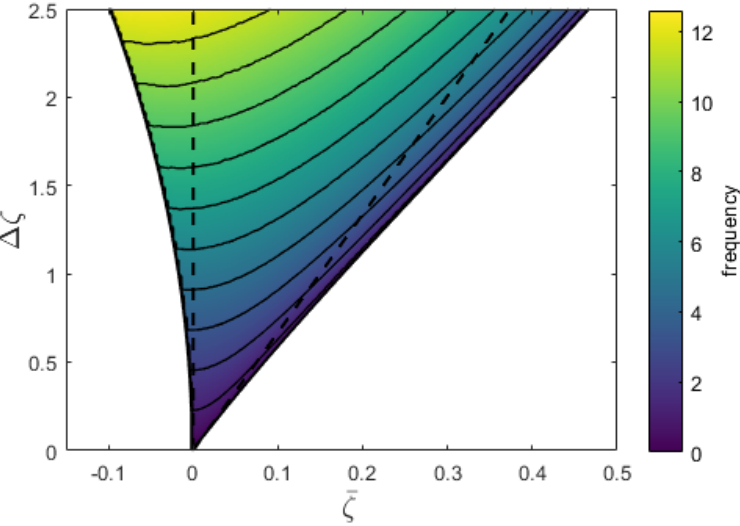}};

        \node[scale=1] at (-0.25,3.25) {a)};
        \node[scale=1] at (5.5,3.25) {b)};
    \end{tikzpicture}
    \caption[The approximate and numerical frequency in the $(\bar\zeta,\,\Delta\zeta)$ parameter plane]{
    a) The approximate and b) the numerically obtained frequency as a function of the average activity, $\bar{\zeta}$, and activity difference across the channel, $\Delta\zeta$. The frequency is non-dimensionalised by the elastic timescale $t_e=\gamma d^2/K$.
    }
    \label{fig:FrequencyApproximation}
\end{figure*}

\end{widetext}

\end{document}